\documentclass{sig-alternate}

\usepackage{graphicx}   
\usepackage{flushend}    
\usepackage{multirow}
\usepackage{multicol}
\usepackage{cite}
\usepackage{subfigure}
\usepackage{float}

\usepackage[table]{xcolor}
\usepackage{epstopdf}
\usepackage{tabularx}
\usepackage{slashbox}
\usepackage{verbatim}
\usepackage{cite}

\usepackage{amsmath}
\usepackage{url}

\def \sys {\textit{Humaine}}
\begin{document}
%
\title{It's the Human that Matters: Accurate User Orientation Estimation for Mobile Computing Applications}

\numberofauthors{2} 
%
\author{
%
%
\alignauthor
Nesma Mohssen\\
       \affaddr{Computer and Systems Eng. Dept.}\\
       \affaddr{Faculty of Engineering, Alexandria Univ., Egypt}\\
       \email{nesma.mohssen.k@alexu.edu.eg}
\alignauthor
Rana Momtaz\\
       \affaddr{Computer and Systems Eng. Dept.}\\
       \affaddr{Faculty of Engineering, Alexandria Univ., Egypt}\\
       \email{rana.momtaz@alexu.edu.eg}
\and
\alignauthor
Heba Aly\\
       \affaddr{Computer and Systems Eng. Dept.}\\
       \affaddr{Faculty of Engineering, Alexandria Univ., Egypt}\\
       \email{heba.aly@alexu.edu.eg}
\alignauthor
Moustafa Youssef\\
       \affaddr{Wireless Research Center}\\
       \affaddr{E-JUST, Egypt}\\
       \email{moustafa.youssef@ejust.edu.eg}
}

\newcommand{\figscale}{0.8}

%


\maketitle

\begin{abstract}
Ubiquity of Internet-connected and sensor-equipped portable devices sparked a new set of mobile computing applications that leverage the proliferating sensing capabilities of smart-phones. For many of these applications, accurate estimation of the user heading, as compared to the phone heading, is of paramount importance. This is of special importance for many crowd-sensing applications, where the phone can be carried in arbitrary positions and orientations relative to the user body. Current state-of-the-art focus mainly on estimating the phone orientation, require the phone to be placed in a particular position, require user intervention, and/or do not work accurately indoors; which limits their ubiquitous usability in different applications.

In this paper we present \sys{}, a novel system to reliably and accurately estimate the user orientation relative to the Earth coordinate system.  
  \sys{} requires no prior-configuration nor user intervention and works accurately indoors and outdoors  for arbitrary cell phone positions and orientations relative to the user body. The system applies statistical analysis techniques to the inertial sensors widely available on today's cell phones to estimate both the phone and user orientation. Implementation of the system on different Android devices with 170 experiments performed at different indoor and outdoor testbeds shows that \sys{} significantly outperforms the state-of-the-art in diverse scenarios, achieving a median accuracy of $15^\circ$ averaged over a wide variety of phone positions. This is $558\%$ better than the-state-of-the-art. The accuracy is bounded by the error in the inertial sensors readings and can be enhanced with more accurate sensors and sensor fusion.
\end{abstract}


%

\section{Introduction}
\label{sec:intro}
\begin{figure}[!t]
\centering
      \includegraphics[width=0.55\linewidth]{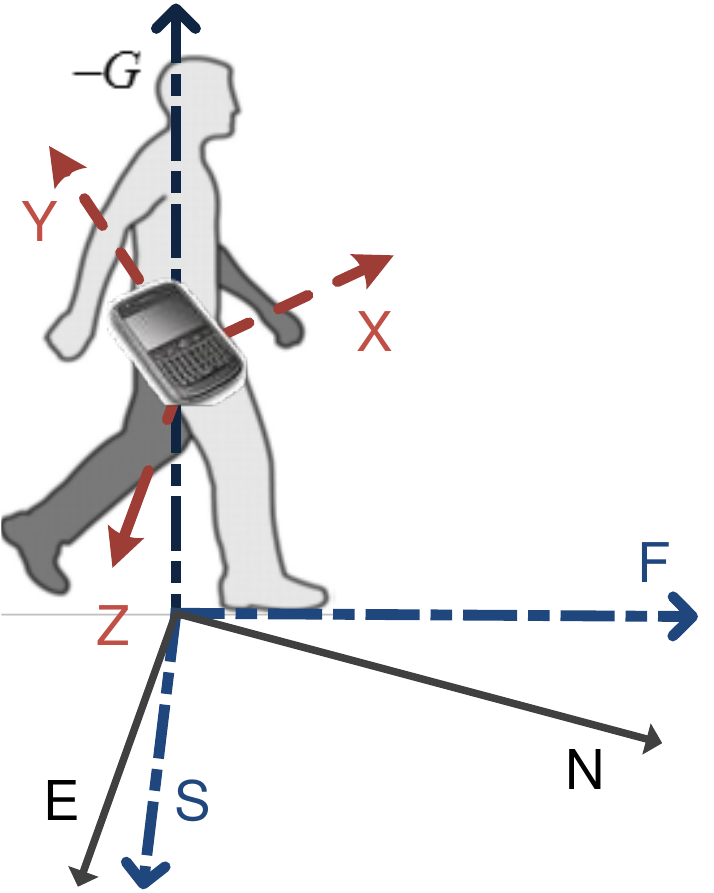}
\caption{The Different coordinate systems: The phone coordinate system $(X, Y, Z)$ is misaligned from the human coordinate system $(F,S,-G)$, which in turn is misaligned with the world coordinate system $(N, E, -G)$. The user plane of motion is the $(F,S)$ plane, which is the plane perpendicular to gravity.}
    \label{fig:coord}
  \end{figure}

Recent advances in ubiquitous computing highlighted the importance of user direction estimation; it enables plentiful ubiquitous and mobile computing applications such as localization \cite{alzantot2012uptime,alzantot2012crowdinside, wang2012no,constandache2010towards,constandache2010compacc}, activity recognition \cite{zhang2012motion}, virtual reality, among others \cite{schall2013mobile,blum2012smartphone,aly_map14}.
Today's smart-phones are equipped with a number of inertial sensors, e.g. accelerometer,  magnetometer, and gyroscope. These sensors provide a measurement for the cell phone orientation relative to the magnetic North. However, users carry cell phones in arbitrary positions and orientations that are typically disoriented with respect to the user (Figure~\ref{fig:coord}). Depending on the phone orientation, rather than the user orientation, can lead to huge errors in many applications. For example, in the popular dead-reckoning localization techniques, e.g. \cite{constandache2010compacc,youssef2010gac}, the user displacement obtained from the inertial sensors is combined with the movement direction to estimate the next user location. Erroneous direction estimation, just based on the returned phone orientation from the OS API, will result in a large error in localization that accumulates quickly with time. This is especially true for the growing field of mobile applications that do not assume a certain placement or orientation of the phone, e.g. in hand or in pocket, such as crowd-sensing applications.

To overcome these problems and obtain the actual user direction, a number of systems have been proposed \cite{omnidirectional,wang2010novel,pca1,pca2,udirect,gluhakdesign,li2012reliable}. For example, \cite{omnidirectional,wang2010novel,pca1,pca2} depend on \textbf{\emph{special external sensors}}  attached to the user's leg or placed in her pants pocket to detect her orientation. Other systems, e.g. \cite{udirect,gluhakdesign,li2012reliable}, use the inertial sensors in standard cell phones to estimate the user direction. However, these solutions either work only in specific phone positions (e.g. pants pocket), are limited to outdoor environments where the phone sensors are not affected by the indoor magnetic noise \cite{udirect,gluhakdesign}, or require special user actions (e.g. \cite{li2012reliable}). These limit their ubiquitous usability.

In this paper, we present \sys{}: a system capable of accurately estimating the user orientation in different environments at arbitrary phone positions and orientations without user intervention. \sys{} starts by fusing the different phone inertial sensors to obtain the phone orientation in the user horizontal plane of motion. To obtain the actual user heading, \sys{} takes advantage of the observation that the direction of motion is the direction that has the maximum acceleration variance. Therefore, it applies the principal component analysis (PCA) on the linear acceleration readings in the user horizontal plane of motion to obtain the final user heading. \sys{} further applies different pre-processing and post-processing steps to reduce the noise effect and remove inherent ambiguity in the direction estimation.

Evaluation of \sys{} on different Android devices with 170 experiments (covering typical homes, a college library, office rooms,  garden, college campus, different wide and narrow streets, among others) shows that it can estimate the user direction with a median error up to $14^\circ$  indoors and $16^\circ$ outdoors for a variety of phone positions. This is better than state-of-the-art direction estimation systems by more than $523\%$ indoors and $594\%$ outdoors.

In summary, our main contributions are three-fold:
\begin{itemize}
\item We present the architecture and details of \sys{}: a system to detect accurate user orientation using standard cell phones at arbitrary positions and orientations in both indoor and outdoor environments.

\item We implement our system on Android-based mobile devices and evaluate its performance as compared to  state-of-the-art systems.

\item We provide a thorough study on the effect of different phone positions (in hand, in pants pocket, in bag, in shirt pocket), phone orientations relative to the user body, and indoor/outdoor effect on the different user direction estimation techniques.
\end{itemize}

The rest of the paper is organized as follows: In Section~\ref{sec:relwork}, we discuss related work. Section~\ref{sec:sys_des} gives the details of the \sys{} system. We provide the implementation and evaluation of the system in Section~\ref{sec:eval}. Finally, Section~\ref{sec:conc} concludes the paper and gives directions for future work.

\section{Related Work}
\label{sec:relwork}
Applications that require the user heading direction estimation, e.g. \cite{constandache2010compacc,youssef2010gac}, usually assume that the phone is oriented with the user. They either depend on the standard cell phone API that uses the magnetometer or they depend on other phone sensors, e.g. the camera \cite{headio}, to obtain more accurate results. However, \textbf{\emph{even with zero error}} in phone direction estimation, using phone-heading estimates can lead to huge errors in case the phone is not aligned with the user direction. Throughout this section, we discuss the different techniques for user direction estimation that have been proposed in literature.

\begin{table*}[!t]
\centering
\resizebox{\textwidth}{!}{\small
\begin{tabularx}{\textwidth}{|l||X|X|X|p{3cm}||X|}
\hline
 \backslashbox{Criteria}{Technique}& Kunze et al.\cite{pca1} & Steinhoff et al.\cite{pca2} & uDirect\cite{udirect,gluhakdesign} &   	Fan Li et al. \cite{li2012reliable} & \textbf{Humaine}\\\hline\hline
Standard Cell-phones & No & No & Yes & Yes & Yes\\\hline
Sensor/phone placement & Pants Pocket & Pants Pocket & Pants Pocket& Pants Pocket and in Hand & \textbf{Anywhere} \\ \hline
Assumptions & User moving forward and not changing direction  & Specific position &  Specific position & Avail. floorplan, known init. head., phone stable rel. to leg mov.& \textbf{None} \\\hline
Evaluation testbed & Outdoors only& Outdoors only& Outdoors  only& Indoors only& \textbf{Outdoors and Indoors}\\\hline
\end{tabularx}
}
\caption{Comparison with the most relevant user orientation detection techniques.}
\label{tab:rw_comp}
\end{table*}
\subsection{Techniques based on Location Estimation}
Frequent logging of user location, e.g. using GPS\cite{gps}, can provide accurate user direction. In this case, the direction is estimated as the direction of the line joining the last two estimated locations. However, this depends on the availability and the accuracy of the localization system. For example, since the GPS accuracy is low in urban areas~\cite{gpsuc}, the direction estimation is not accurate. In addition, it completely fails in indoor environments and has significant \textbf{latency and energy-consumption} as we quantify in Section~\ref{sec:eval}.

\subsection{Techniques that use Special Sensors}
To overcome this limitation, researchers employed other sensors (e.g. wearable cameras, shoe-mounted sensors, and/or inertial head trackers) to provide accurate direction estimation\cite{kourogi2003method, kourogi2003wearable,kourogi2003personal, wang2010novel,omnidirectional,pca1,pca2}. 
In \cite{pca1,pca2,kourogi2003personal}, a \emph{\textbf{special acceleration sensor}}, e.g. the MTx motion sensor \cite{pca1} is used to obtain the motion axis that is parallel to the movement direction, but it fails to determine the forward direction itself, leading to a ``180$^\circ$ ambiguity problem''. \cite{pca1} addressed the 180$^\circ$ ambiguity problem using integration of the acceleration signal in the global frame, which does not yield a robust estimate~\cite{pca2}.
To detect the forward direction, \cite{pca2} leverages the rotational motion of the sensor before foot impact. However, using the foot impact to solve the 180$^\circ$ ambiguity leads to a problem when applying the technique indoors as we show in Section~\ref{sec:eval}.

In \cite{kourogi2003personal}, authors used inertial sensors, a wearable camera, and an inertial head tracker. The forward direction is determined by testing whether the slope of vertical acceleration at the peak for forward acceleration is increasing. This algorithm requires sensors to be attached to the torso for correct detection of the acceleration patterns, limiting its applicability for different positions. Also, using a wearable camera imposes further limitations on the applicability of the technique and the environment (e.g. lightning).

Different from all these techniques, \sys{} provides a robust and accurate heading estimation indoors and outdoors using inertial sensors available on commodity cell phones (which usually have lower quality compared to the special sensors used in these techniques) with no restriction on the phone position.

\subsection{Techniques based on Cell Phone Sensors}
Recently, researchers have focused on using standard cell phones sensors to detect the user heading direction.
In \cite{li2012reliable}, the system \textbf{assumes that initially the cell phone is in the pants pocket and the heading offset is known}. The system leverages the periodicity of the leg movement during walking to identify a point during each step where the relative orientation of the phone to the user's body is the same as in the initial standing state. 
The system uses a particle filter to mitigate the magnetic field noise effect.
  However, this particle filter requires a map of the building, which may not be ubiquitously available especially for crowd-sensing applications.

The uDirect system \cite{udirect,gluhakdesign} employs a similar technique to estimate the user direction; They identify a point in the middle between the user's detected heel strike and toe-off moments as the point where the device orientation is close to the device orientation in the standing mode. 
These systems, however, require a model for the acceleration pattern within a step for each phone position; A model for the phone placed in the pants pocket was presented in~\cite{udirect,gluhakdesign}. Deriving the model for a new position is not straightforward and the acceleration pattern for other positions may not be as clear as in the case of the presented pants pocket. 
 Furthermore, the magnetic field noise, which affects the \emph{oriented/rotated} acceleration pattern used in heading estimation, degrades uDirect accuracy indoors as we quantify in Section~\ref{sec:eval}.

\subsection{Summary}
Table~\ref{tab:rw_comp} summarizes the differences between \sys{} and the most relevant state-of-the-art. The current state-of-the-art either require special external sensors, work in a specific phone position (e.g. in pants pocket) or orientation (e.g. oriented phone with the user), require user intervention, and/or work only outdoors. 

\sys{}, on the other hand, depends on available inertial sensors in standard cell phones. It is applicable indoors and outdoors for arbitrary phone positions and orientations without any intervention from the user.

\begin{table}[!t]
\centering
\begin{tabularx}{\linewidth}{|c|X|}\hline
Symbol & Description \\\hline\hline
$N, E, -G$ & The world axes (Figure~\ref{fig:coord}): $N$ points to North, $E$ points to East, and $-G$ is opposite to gravity.\\
$F,S,-G$ & The axes of the user coordinate system (Figure~\ref{fig:coord}): $F$ points in the forward user motion direction, $S$ points toward the side, and $-G$ is opposite to gravity.\\
$X,Y,Z$ & The axes of the phone coordinate system (Figure~\ref{fig:pln_trns}): $X$ points towards the phone side, $Y$ points toward the phone head, and $Z$ is perpendicular to phone screen.\\
$\alpha$  & The phone pitch angle (Figure~\ref{fig:pln_trns}).\\
$\beta$  & The phone roll angle (Figure~\ref{fig:pln_trns}).\\
$\gamma$  & The phone yaw angle relative to North (phone azimuth, Figure~\ref{fig:pln_trns}).\\
$\theta_u$ & The user orientation angle relative to North.\\
$R_A(b)$ & Rotation matrix of angle $b$ around axis $A$.\\
$g$ & The Earth's gravity acceleration ($9.80665 m/s^2$).\\
$a_t$ & Total acceleration affecting the phone in the phone coord. system (Figure~\ref{fig:grav_lin}).\\
$\ell$ &  Linear acceleration in the phone coordinate system (Figure~\ref{fig:grav_lin}).\\
$\ell_u$ & $(\ell_F,\ell_S)$, the linear acceleration in the user coordinate system.\\
$\psi$& The dip angle of the magnetic field measured downwards from the horizontal plane \cite{ronald1998magnetic} for a perfectly oriented phone, i.e. pointing to North.\\
$G_m$ & $[g_x, g_y, g_z]^t$, gravity acceleration components in the phone coordinate system (Figure~\ref{fig:grav_lin}).\\
$M$ & $[m_x, m_y, m_z]^t$, the measured magnetic field in the phone coordinate system.\\
$\delta$ & Noise filter smoothing factor.\\
$\omega$ & PCA window size.\\
\hline
\end{tabularx}
\caption{Summary of the symbols used.}
\label{tab:symbols}
\end{table}

\section{The Humaine System} \label{sec:sys_des}
In this section, we provide the details of the different components of the system; covering the sensor readings preprocessing, the acceleration transformation to the user's plane, detection of the user's motion axis, and finally detection of the user's orientation relative to North. We start by defining the coordinate systems used in the paper, followed by an overview of the system architecture, and finally the details of each module.

\begin{figure*}[!t]
\centering
\includegraphics[width=0.95\linewidth]{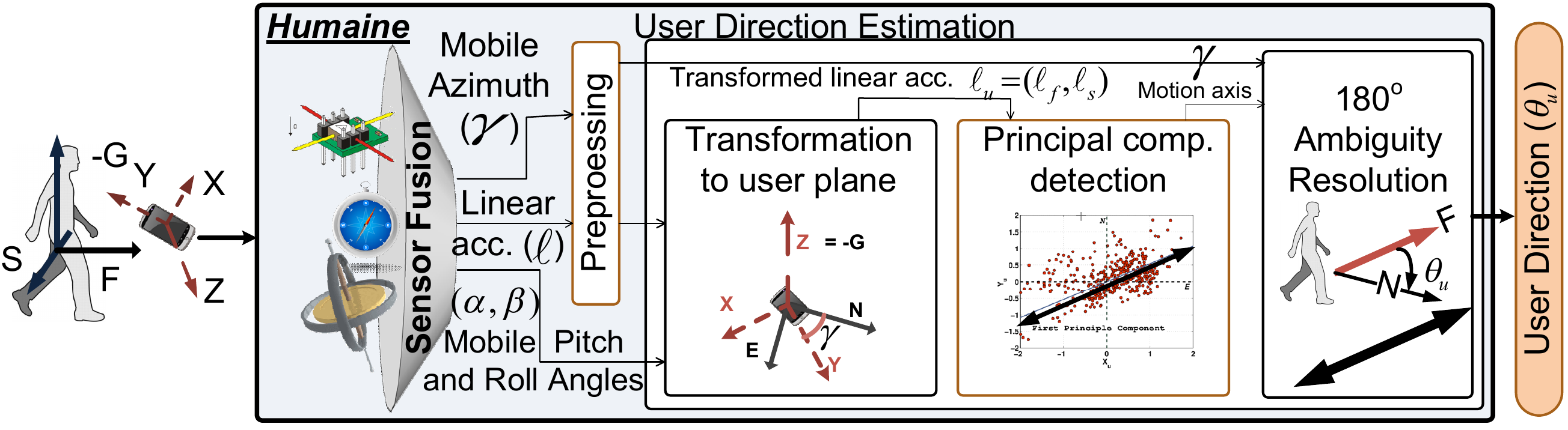}
\caption{\sys{} system architecture --- The system takes the raw sensor readings from the user's cell phone and gives the user's direction relative to North.}
\label{fig:sys_arch}
\end{figure*}

\subsection{Coordinate Systems}

Figure~\ref{fig:coord} shows the different coordinate systems used in the paper. The world coordinate system $(N, E, -G)$ is defined by North ($N$), East ($E$), and the Earth gravity ($-G$). We refer to the phone local coordinate system as $(X, Y, Z)$. The user plane of motion is perpendicular to gravity ($-G$) and we are interested in the user forward direction ($F$). Therefore, the user coordinate system is defined as $(F, S, -G)$ where $S$ points toward the right side of the user's forward direction ($F$).
Table~\ref{tab:symbols} summarizes all symbols used in this section.

\subsection{System Overview}

Figure \ref{fig:sys_arch} shows the system architecture. The system detects the orientation, relative to North, of a user that carries a cell phone with her in an arbitrary position using the available cell phone's inertial sensors.

The Sensor Fusion Module fuses the different inertial sensors to obtain the linear acceleration, the phone azimuth direction, and the phone rotation angles relative to the world.

The Preprocessing Module filters the linear acceleration readings to reduce the noise.

The User Direction Estimation Module transforms the linear acceleration readings to the user plane of motion and estimates the user motion axis as the direction with the maximum variance. Finally, the Ambiguity Resolution Module disambiguates the final user's heading direction along the motion axis.

\begin{figure}[!t]
\centering
\includegraphics[width=0.7\linewidth]{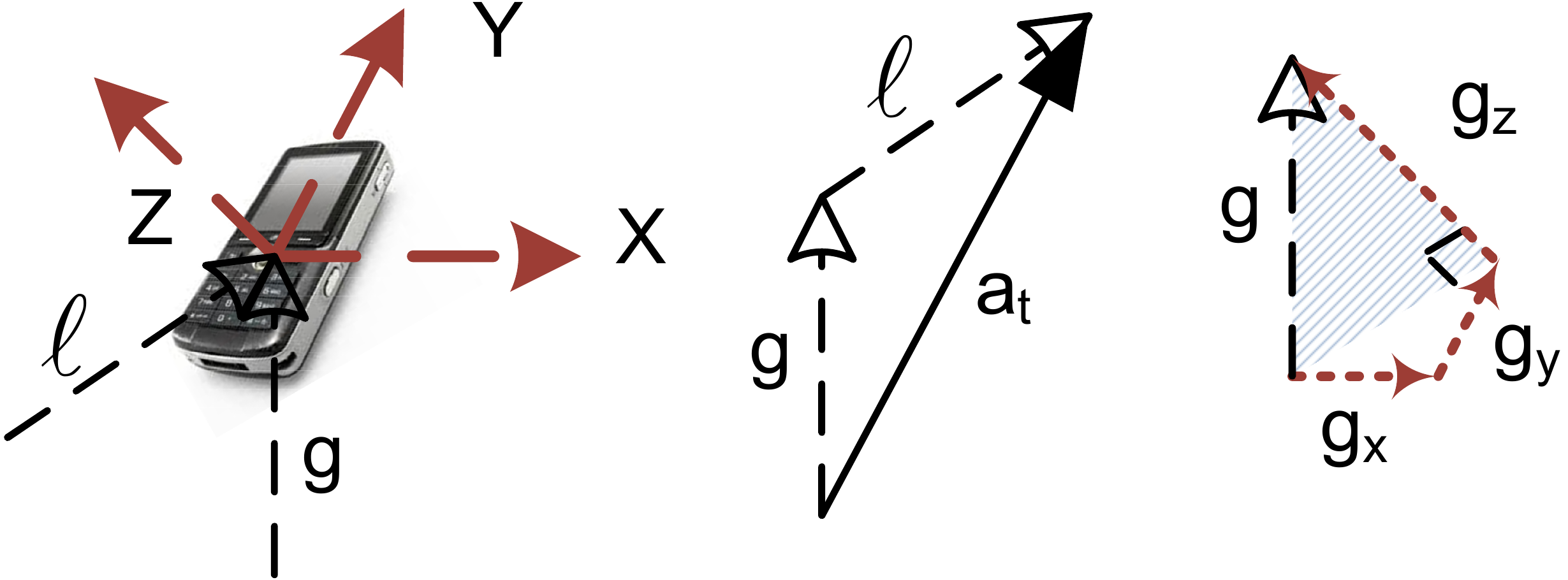}
\caption{The total force ($a_t$) applied to the phone at any time instance is the sum of the gravity acc. ($g$) and the linear acc. ($\ell$). The gravity acc. components in the phone coordinate system are $(g_X, g_Y, g_Z)$.}
\label{fig:grav_lin}
\end{figure}

\begin{figure}
  \centering
  \includegraphics[width=1.1\linewidth]{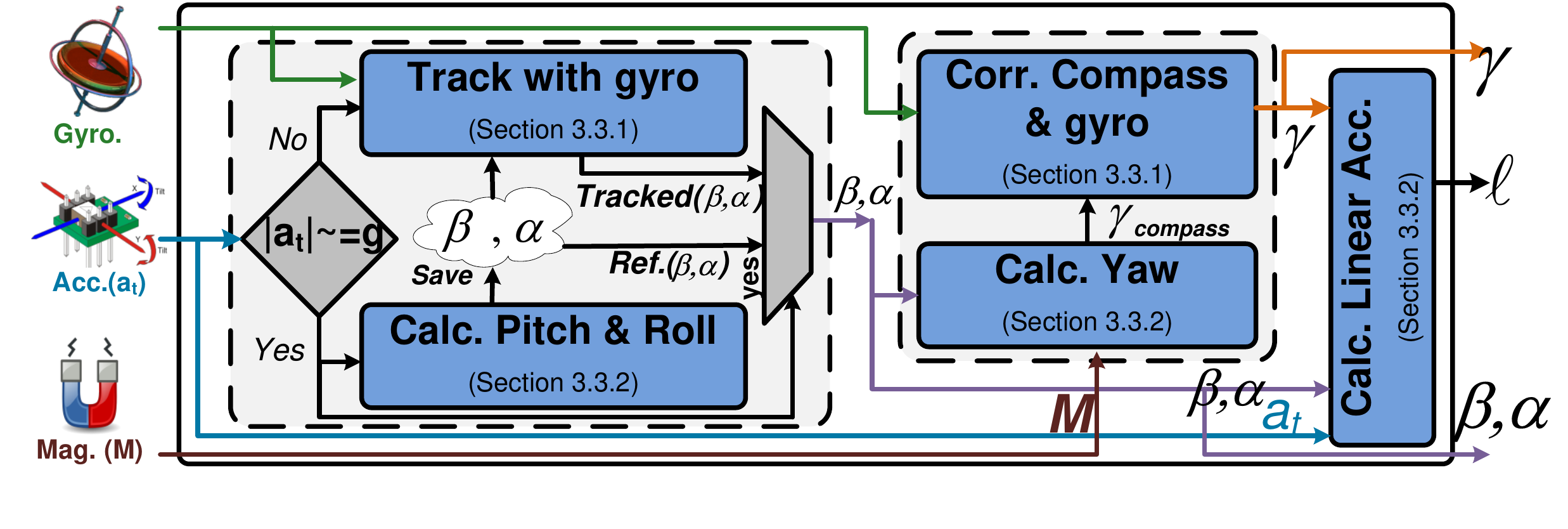}\\
  \caption{Sensor fusion: The inertial sensors are combined to obtain an accurate phone orientation in the presence of sensors noise and drift.}
  \label{fig:fusion}
\end{figure}

\begin{figure}[!t]
\centering
      \includegraphics[width=1\linewidth]{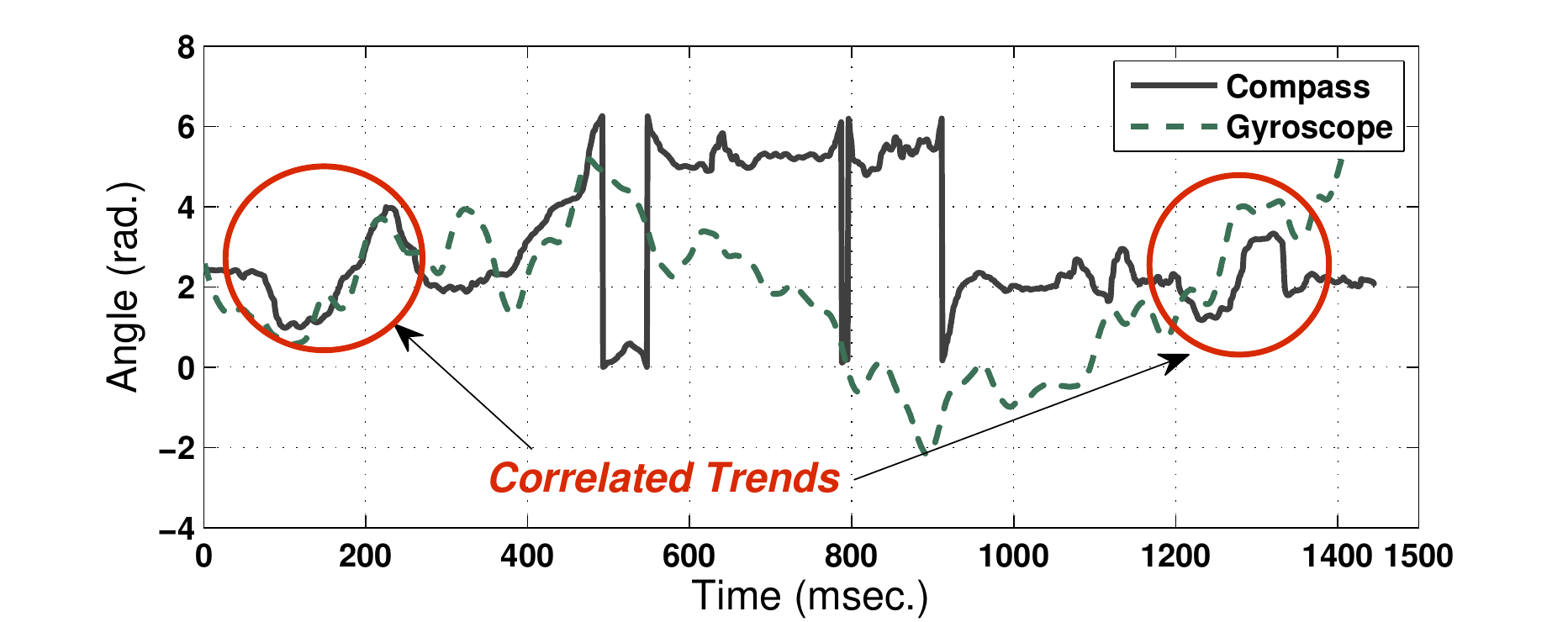}
\caption{Example showing leveraging the correlation between the compass and gyroscope sensors to determine when the compass readings are reliable.}
    \label{fig:comp_gyro}
  \end{figure}

\subsection{Sensor Fusion}
\label{sec:fusion}
The system collects raw sensor information from inertial sensors available in the cell phone. In particular, we collect the 3D acceleration, the 3D magnetic field from the magnetometer, and the relative rotation angle from the gyroscope.
Sensor fusion is an important step to reduce noise and estimate the quantities of interest. Figure~\ref{fig:fusion} shows an overview of our sensor fusion module.

\subsubsection{Gravity Acceleration Estimation}
\label{sec:g_est}
The accelerometer returns the acceleration force applied to the phone which consists of two components: the gravity and linear acceleration applied to the phone due to its motion (Figure~\ref{fig:grav_lin}). Since, we are only interested in the device's acceleration due to motion, we need to remove the gravity component.  To do so, \sys{} opportunistically uses the instances when the magnitude of the acceleration vector approximately equals $g (9.80665m/s^2)$ as the gravity reference (Figure~\ref{fig:fusion}). The intuition is that, at these instances, there are no other forces applied to the phone. Otherwise, \sys{} uses the gyroscope to track the gravity vector angles as it moves due to the user movement. Then, the gravity vector is factored out to obtain the linear acceleration in the phone coordinate system as described below in Section~\ref{sec:phone2user}.

However, since the gyroscope sensor suffers from drift, i.e. accumulation of error with time, we fuse it with the compass readings, which has long term stability but suffers from short term magnetic noise, to obtain reliable orientation readings. To determine the points in time when the compass reading is reliable, we depend on the the correlation between the compass and the gyroscope readings (Figure~\ref{fig:comp_gyro}). When both sensors exhibit a similar pattern, we declare that the compass reading is accurate and can be used to correct the gyroscope drift \cite{wang2012no}.

\subsubsection{Phone Orientation Estimation}
\label{sec:phone2user}
\begin{figure}[!t]
\centering
\includegraphics[width=0.5\linewidth]{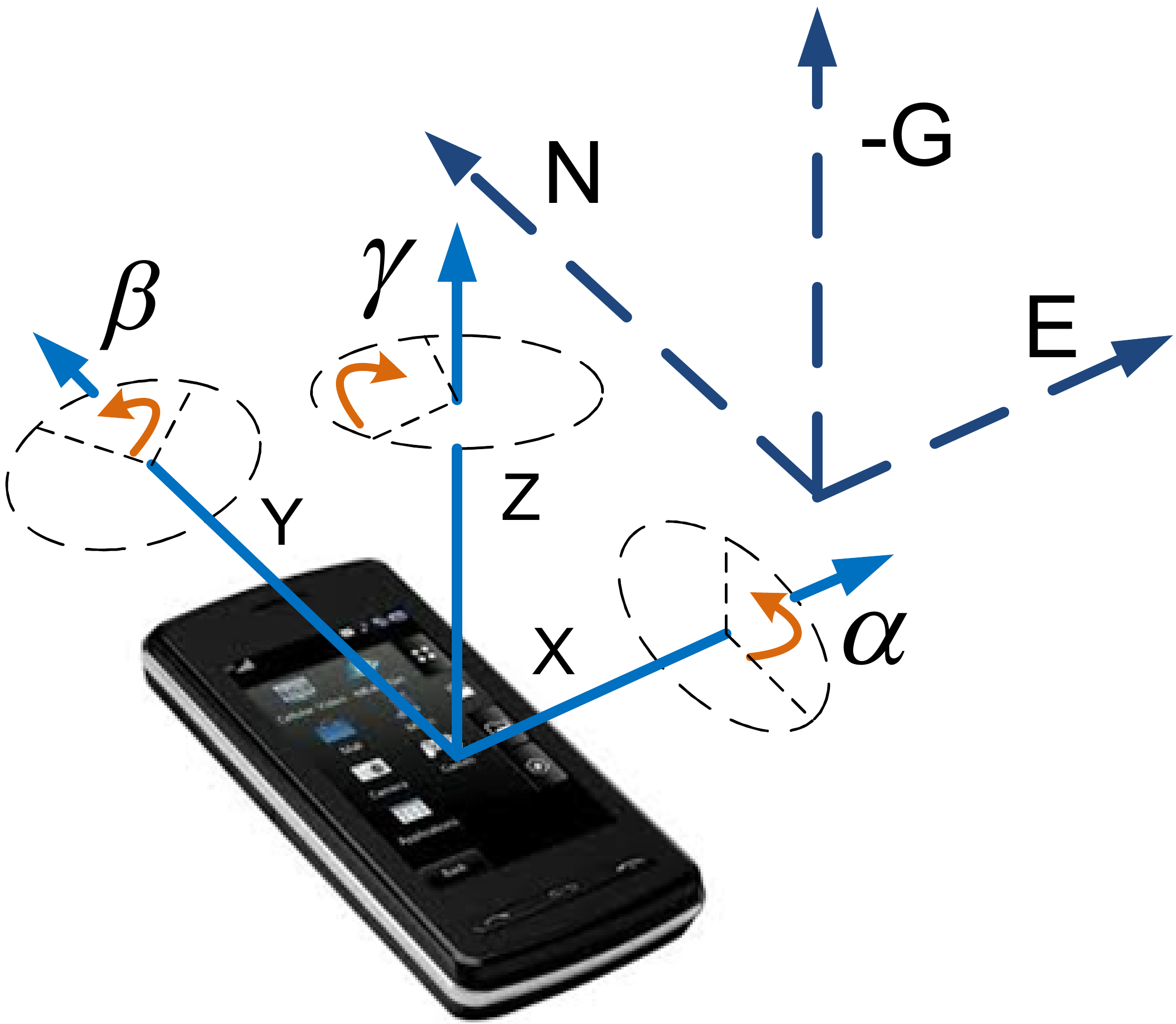}
\caption{Different phone orientation angles: pitch ($\alpha$), roll ($\beta$), and yaw ($\gamma$). $\gamma$ is the phone orientation angle relative to North (azimuth angle).}
\label{fig:pln_trns}
\end{figure}
\begin{figure}[!t]
\centering
      \includegraphics[width=0.5\linewidth]{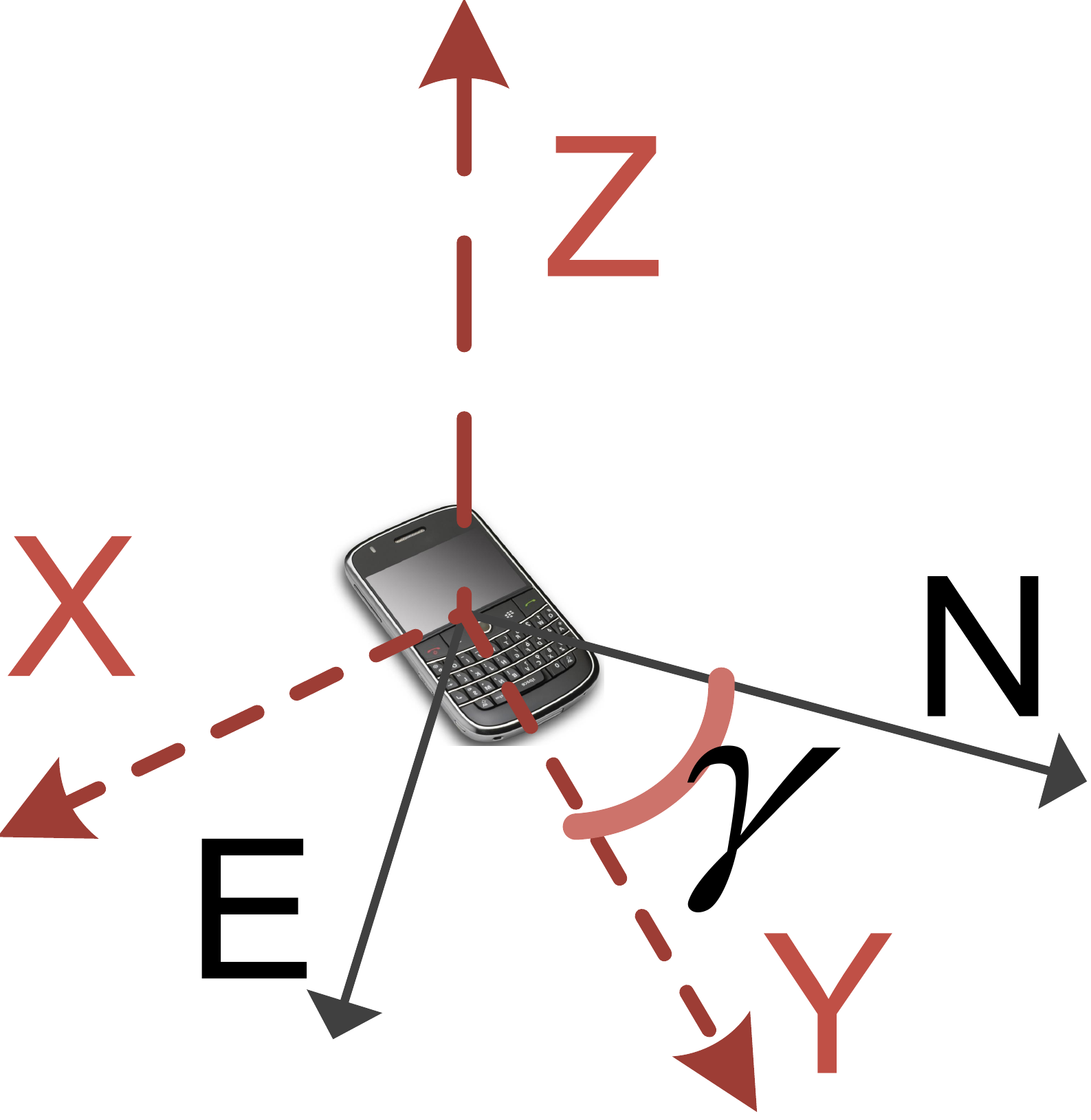}
\caption{The phone orientation after applying the rotation matrices $R(-\beta)R(-\alpha)$. The phone $Z$-axis becomes aligned with $-G$ and the phone rests in the horizontal plane of motion making an angle $\gamma$ with North (phone orientation angle).}
    \label{fig:phone_to_user}
  \end{figure}

Users carry cell phones in different positions (e.g. pants' pocket, shirt's pocket, in bag, or in hand). Therefore, the phone can have an arbitrary orientation relative to the user's plane of motion. To estimate the phone orientation, we depend on the estimated gravity acceleration (Section~\ref{sec:g_est}). For ease of illustration, we first use Euler angles and rotation matrices to explain how \sys{} estimates the phone orientation angles around the three main axes; i.e. yaw, pitch, and roll (Figure~\ref{fig:pln_trns}). Then, we explain our actual implementation.

Let $G_m=[g_x, g_y, g_z]^t$ be the gravity components, of the estimated gravity acceleration affecting the phone in its arbitrary location. $G_m$ can be written as:
\begin{equation}
\begin{matrix}
\begin{bmatrix}
g_x\\
g_y\\
g_z
\end{bmatrix} & = & R_X(\alpha)R_Y(\beta)R_Z(\gamma)\begin{bmatrix}
0\\
0\\
g
\end{bmatrix}
\end{matrix}
\label{eq:main}
\end{equation}
where $R_A(b)$ is the rotation matrix representing a rotation with angle $b$ around axis $A$. In particular:
\begin{equation}
\begin{matrix}
R_X(\alpha)& =&
\begin{bmatrix}
1 & 0 & 0\\
0 & \cos(\alpha) & \sin(\alpha)\\
0 & -\sin(\alpha) & \cos(\alpha)
\end{bmatrix}
\end{matrix}
\label{eq:rx}
\end{equation}
\begin{equation}
\begin{matrix}
R_Y(\beta)& =&
\begin{bmatrix}
\cos(\beta) & 0 & -\sin(\beta)\\
0 & 1 & 0\\
\sin(\beta) & 0 & \cos(\beta)
\end{bmatrix}
\end{matrix}
\label{eq:ry}
\end{equation}
\begin{equation}
\begin{matrix}
R_Z(\gamma)& =&
\begin{bmatrix}
\cos(\gamma) & \sin(\gamma) & 0\\
-\sin(\gamma) & \cos(\gamma) & 0\\
0 & 0 & 1\\
\end{bmatrix}
\end{matrix}
\label{eq:rz}
\end{equation}
To obtain the pitch ($\alpha$) and roll ($\beta$) angles, 
 we multiply both sides of Equation~\ref{eq:main} by $R^{-1}_Y(\beta) R^{-1}_X(\alpha)$ and noting that $R^{-1}_A(b)= R_A(-b)$, we obtain:
\begin{equation}
R_Y(-\beta) R_X(-\alpha)G_m= R_Z(\gamma)\begin{bmatrix}
0\\
0\\
g
\end{bmatrix}
\label{eq:approach}
\end{equation}
Substituting from equations~\ref{eq:rx}, \ref{eq:ry}, and \ref{eq:rz} we obtain:
\begin{equation}
\begin{matrix}
\begin{bmatrix}
\cos(\beta) & \sin(\beta) \sin(\alpha)& \sin(\beta) \cos(\alpha)\\
0 & \cos(\alpha) & -\sin(\alpha) \\
-\sin(\beta) & \cos(\beta)\sin(\alpha) & \cos(\beta)\cos(\alpha)\\
\end{bmatrix}
\begin{bmatrix}
g_x\\
g_y\\
g_z
\end{bmatrix}
&
=&
\begin{bmatrix}
0\\
0\\
g
\end{bmatrix}
\end{matrix}
\label{eq:alpha_full}
\end{equation}
From the second and first rows of Equation~\ref{eq:alpha_full} we get:
\begin{equation}
\tan(\alpha)= \frac{g_y}{g_z}
\label{eq:alpha}
\end{equation}
\begin{equation}
\tan(\beta)= \frac{-g_x}{g_y \sin(\alpha)+ g_z \cos(\alpha)}
\label{eq:beta}
\end{equation}
To obtain the yaw angle ($\gamma$), which is the phone orientation angle relative to North (Figure~\ref{fig:phone_to_user}), we leverage the magnetometer signal. In particular, let $M= [m_x, m_y, m_z]^t$ be the magnetic field at the arbitrary phone orientation.  Using a similar approach to Equation~\ref{eq:approach}, we get
\begin{multline}
\begin{bmatrix}
m_x\cos(\beta)+ m_y \sin(\beta) \sin(\alpha)+ m_z \sin(\beta) \cos(\alpha)\\
m_y \cos(\alpha) - m_z \sin(\alpha) \\
-m_x \sin(\beta) + m_y \cos(\beta)\sin(\alpha) + m_z\cos(\beta)\cos(\alpha)\\
\end{bmatrix}
\\
=
\begin{bmatrix}
E \cos(\psi) \sin(\gamma)\\
-E \cos(\psi) \cos(\gamma)\\
E \sin(\psi)
\end{bmatrix}
\label{eq:gamma_full}
\end{multline}
where $E$ and $\psi$ are the \textbf{\emph{unknown}} Earth magnetic field strength and the dip angle of the field measured downwards from horizontal \cite{ronald1998magnetic} for a perfectly oriented phone, i.e. pointing to North. From the first and second rows of Equation~\ref{eq:gamma_full} we get:
\begin{equation}
  \tan(\gamma)= \frac{m_x\cos(\beta)+ m_y \sin(\beta) \sin(\alpha)+ m_z \sin(\beta) \cos(\alpha)}{m_z \sin(\alpha)- m_y \cos(\alpha)}
  \label{eq:gamma}
\end{equation}
Equations \ref{eq:alpha}, \ref{eq:beta}, and \ref{eq:gamma} are solved taken into account the sign of the gravity acceleration and magnetometer components to obtain the correct quadrant for the respective angles.\\\\
\begin{figure}[!t]
\centering
\includegraphics[width=0.5\linewidth]{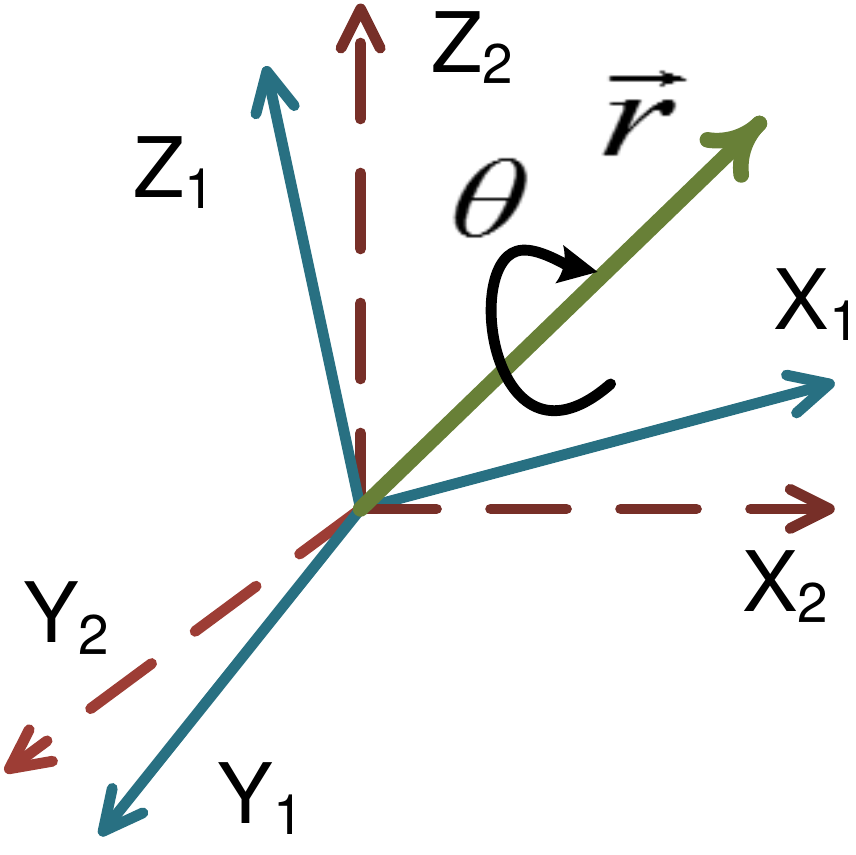}
\caption{Transforming between two coordinate systems can be performed using three rotations around the three orthogonal axes by the corresponding Euler angles. This is equivalent to a single rotation around axis $\vec{r}$ with an angle $\theta$.}
\label{fig:rv_vis}
\end{figure}
\textbf{\textit{Actual Implementation}}

We use unit quaternions to represent the rotation operation in \sys{} rather than rotation matrices. This representation evolves from Euler's rotation theorem\cite{goldstein1980classical}, which implies that any rotation or sequence of rotations of a rigid body in a three-dimensional space is equivalent to a pure rotation about a single fixed axis $r=[r_x,r_y,r_z]^t, r_x^2+r_y^2+r_z^2=1$ by an angle $\theta$ (Figure~\ref{fig:rv_vis}). Quaternions give a simple way to encode the rotation operation in 4 parameters, compared to 9 parameters in case of rotation matrices. In addition, quaternions suffer from fewer mathematical rounding defects and are not subject to the gimbal lock problem \cite{hamilton1866elements,goldstein1980classical}.

The relation between the pitch, roll, and yaw angles obtained in the previous section and the 4D quaternion vector is as follows:
\begin{multline}
q=
\begin{bmatrix}
\cos(\frac{\theta}{2})\\
\sin(\frac{\theta}{2}) r_x\\
\sin(\frac{\theta}{2}) r_y\\
\sin(\frac{\theta}{2}) r_z
\end{bmatrix}
\\=
\begin{bmatrix}
\cos(\frac{\alpha}{2})\cos(\frac{\beta}{2})\cos(\frac{\gamma}{2})+\sin(\frac{\alpha}{2})\sin(\frac{\beta}{2})\sin(\frac{\gamma}{2})\\
\cos(\frac{\alpha}{2})\sin(\frac{\beta}{2})\cos(\frac{\gamma}{2})- \sin(\frac{\alpha}{2})\cos(\frac{\beta}{2})\sin(\frac{\gamma}{2})\\
\sin(\frac{\alpha}{2})\cos(\frac{\beta}{2})\cos(\frac{\gamma}{2})+ \cos(\frac{\alpha}{2})\sin(\frac{\beta}{2})\sin(\frac{\gamma}{2})\\
\cos(\frac{\alpha}{2})\cos(\frac{\beta}{2})\sin(\frac{\gamma}{2})-\sin(\frac{\alpha}{2})\sin(\frac{\beta}{2})\cos(\frac{\gamma}{2})
\end{bmatrix}
\label{eq:quat}
\end{multline}

\subsection{Preprocessing}
\label{sec:preprocessing}
\sys{} estimates the user direction based on processing the linear acceleration vector from raw sensor measurements; These measurements are sensitive to abrupt changes in the cell phone, e.g. due to shaking. We apply a low-pass filter on the linear acceleration to reduce the effect of the noisy measurements with the following equation:
\[  s(i) = s(i-1) + \delta . (r(i) - s(i-1)), i>0 \]
Where $s(i)$ is the $i^{\textrm{th}}$ smoothed linear acceleration signals, $r(i)$ is the $i^{\textrm{th}}$ raw linear acceleration sample, and $\delta$ is the smoothing factor. A smoothing factor of 0.25 gave us the best performance in our experiments.

\subsection{Obtaining the User Motion Direction}
\label{sec:pca}
\begin{figure}[!t]
\centering
\includegraphics[width=0.9\linewidth]{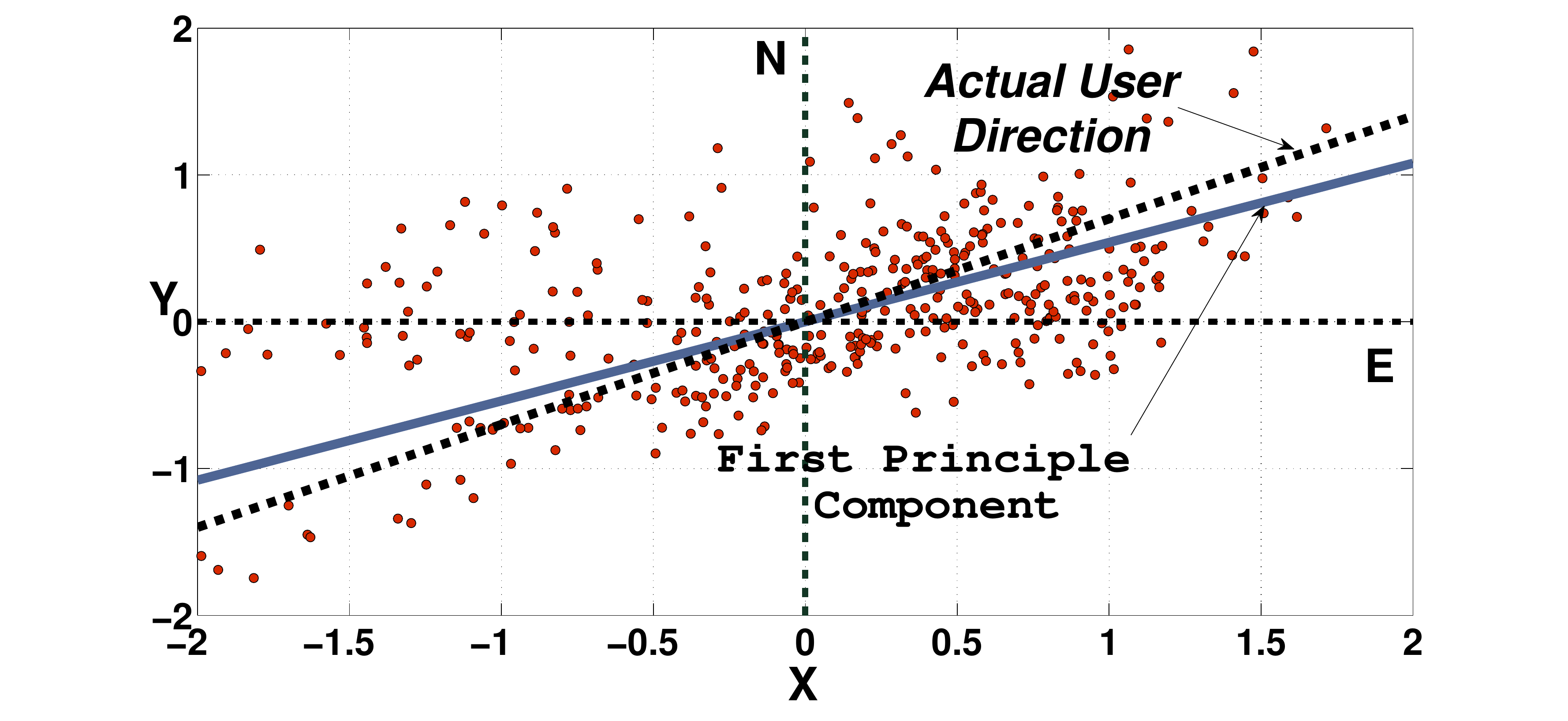}
\caption{PCA applied to the linear acceleration samples (red points) in the user plane after orienting the phone.  The figure shows the estimated as well as the actual user direction.}
\label{fig:usr_dir}
\end{figure}

\begin{figure}[!t]
\centering
\includegraphics[width=0.9\linewidth]{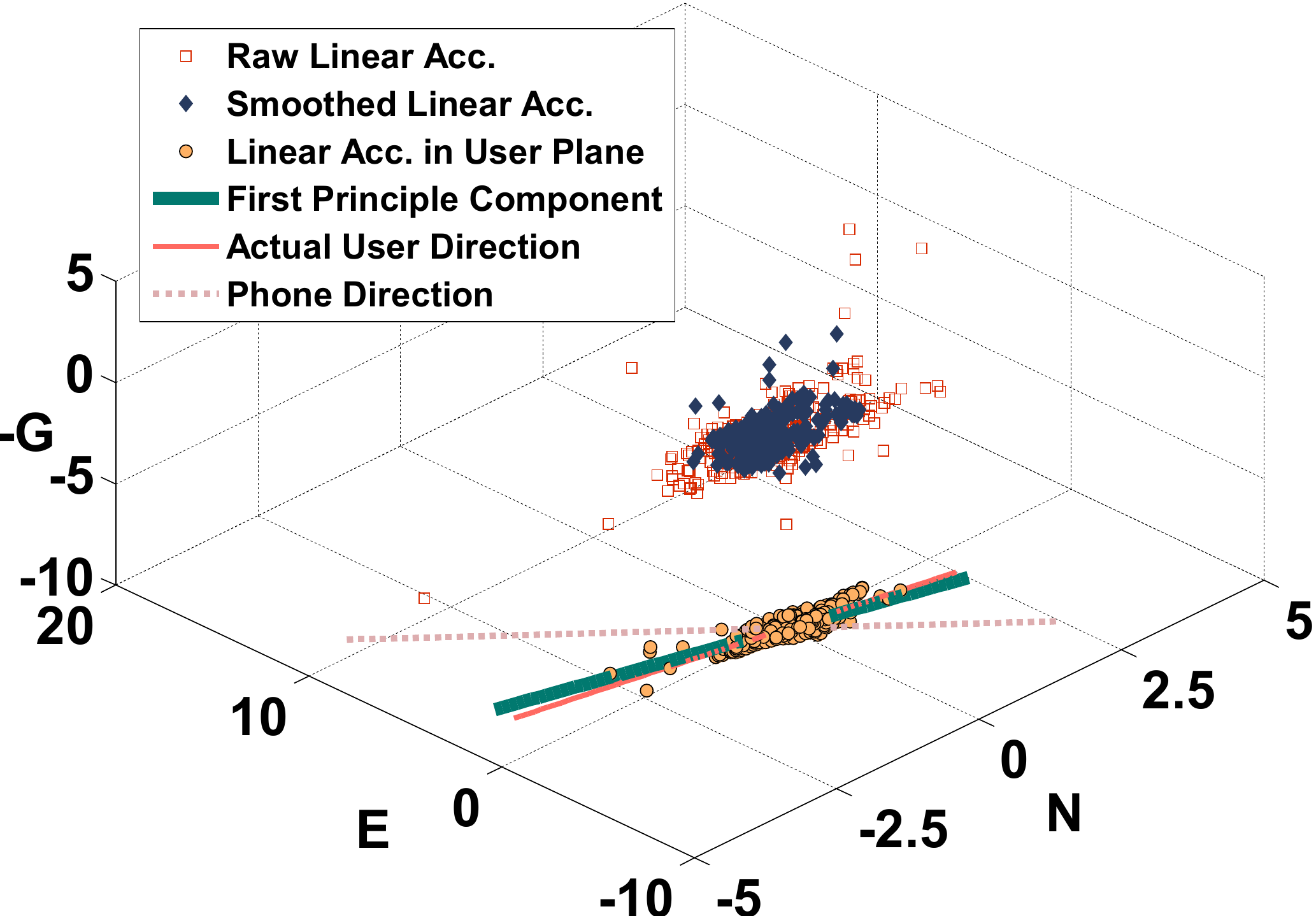}
\caption{The figure shows the linear acceleration at different processing steps by \sys{} to get the motion axis: raw acceleration, de-noised, and transformed linear acceleration.}
\label{fig:pca}
\end{figure}

To obtain the user motion direction, we transform the linear acceleration ($\ell$), which is the acceleration due to all forces applied to the phone except gravity, from the cell phone coordinate system to the user coordinate system to obtain $\ell_u$. This is achieved by applying the quaternions to the linear acceleration vector ($\ell$):
\[ \ell_u = q . \ell . q^* \]
where: $q^*$ is the conjugate of the quaternion $q$.

Once the phone linear acceleration is transformed to the user coordinate system, what remains is to obtain the user direction relative to North. Figure~\ref{fig:usr_dir} shows that even though the phone has been oriented to North by applying the rotation angles, the linear acceleration samples are not aligned with North, but rather with the user direction. \sys{}  exploits this observation to detect the user direction as the direction that has the maximum variance. To obtain this direction, we consider only the two horizontal components of the transformed linear acceleration ($\ell_u=(\ell_F,\ell_S)$) that lie in the (F,S) plane, which is the user plane of motion. We then apply PCA on a window that contains values of ($\ell_F,\ell_S$) components of the transformed linear acceleration ($\ell_u$). The first principal component is the direction of maximum variance. Figure~\ref{fig:pca} shows the acceleration signal at different stages of processing by \sys{}. We found that a window size ($\omega$) of 3 seconds balances accuracy and latency.
 
\subsection{Resolving Ambiguity}

The obtained direction through PCA cannot differentiate between the actual user direction and its opposite, i.e. $\theta_u$ and $\theta_u+180$. To resolve this ambiguity, we use the phone orientation angle ($\gamma$) as a hint and choose the PCA output direction that is closest to $\gamma$. This will work as long as the difference between the y-axis of the phone and the forward direction of the user is within $\pm$90$^\circ$, which is the typical case as we see in Section~\ref{sec:eval}.

\section{Evaluation}
\label{sec:eval}
We implemented \sys{} on different Android devices including Samsung Galaxy Nexus S, Samsung Galaxy Tab 10.1, and an Asus Nexus 7 tablet. Raw sensor measurements were obtained through the Android API with sampling rate 50 Hz. 
To evaluate the system applicability under different magnetic field characteristics, we used 17 testbeds including different rooms at a typical home environment, a college library, office rooms,  garden, college campus, different wide and narrow streets, among others. A {\it \bfseries total of 170 experiments with more than 7000 user direction estimates} were performed by 10 users while placing the phones at different positions including pants pocket, shirt pocket, in bag, and in hand; as well as covering different phone orientations relative to the user direction of motion.
To obtain ground truth, we marked the user path on the ground and used the path orientation from Google Maps as the ground-truth.

For the rest of the section, we compare the performance of the end-to-end system, in typical indoor and outdoor testbeds under different phone positions, with the Android API (as the current widely deployable system), uDirect system \cite{udirect,gluhakdesign} (closest related work), and GPS. The metrics used are absolute angle estimation error, latency, and power consumption.

\subsection{Comparison Systems}
\label{sec:comparison}
\begin{table*}[!t]
\centering
\resizebox{1.05\textwidth}{!}{\small
\begin{tabular}{|c||l|l|l||l|l|l||l|l|l||l|l|l|}\hline
\multirow{3}{*}{Technique} & \multicolumn{12}{c|}{User Heading Estimation Error \emph{Indoors} (degrees)}\\ \cline{2-13}
& \multicolumn{3}{c||}{Pants pocket} &\multicolumn{3}{c||}{Hand held} &  \multicolumn{3}{c||}{Shirt pocket} & \multicolumn{3}{c|}{Bag}\\\cline{2-13}
& 50\% & 75\% & Max.  & 50\% & 75\% & Max. & 50\% & 75\% & Max. & 50\% & 75\% & Max. \\ \hline \hline
\sys{} & 21 & 31 & 55 & 12 & 19 & 30 & 9 & 16 & 45  &17&24&40
 \\ \hline
\multirow{2}{*}{uDirect}&  30 & 50 & 145 & 102 & 141 & 180& 103 & 142 & 180  &90&137&175\\
&(42.9\%) &	(61.3\%) &	(163.6\%) &(750.0\%) &	(642.1\%) &	(500.0\%) &(1044.4\%) &	(787.5\%) &	(300.0\%) &(429.4\%) &(470.8\%) & (337.5\%)\\
  \hline
\multirow{2}{*}{Android API} & 58 & 63 & 180 & 52 & 62 & 100 &  125 & 165 & 180  & 42 & 130 & 180\\
&	(176.2\%) &	(103.2\%) &	(227.3\%)&	(333.3\%) &	(226.3\%) &	(233.3\%) &		(1288.9\%) &	(931.3\%) &	(300.0\%) &(147.1\%) &(441.7\%) &(350.0\%)  \\ \hline

\hline\end{tabular}
}
\caption{Comparison between \sys{}, uDirect\cite{udirect,gluhakdesign}, and Android API in \emph{\bfseries indoor} environments. Note that GPS does not work indoors. Percentages degradation are calculated relative to \sys{}.}
\label{tab:comp_in}
\end{table*}

\begin{table*}
\centering
\resizebox{1.05\textwidth}{!}{\small
\begin{tabular}{|c||l|l|l||l|l|l||l|l|l||l|l|l|}\hline
\multirow{3}{*}{Technique} & \multicolumn{12}{c|}{User Heading Estimation Error \emph{Outdoors} (degrees)} \\\cline{2-13}
&  \multicolumn{3}{c||}{Pants pocket} &\multicolumn{3}{c||}{Hand held} &  \multicolumn{3}{c||}{Shirt pocket}  & \multicolumn{3}{c|}{Bag}\\\cline{2-13}
&50\% & 75\% & Max. & 50\% & 75\% & Max. & 50\% & 75\% & Max.& 50\% & 75\% & Max. \\ \hline \hline
\sys{} &  24 & 30 & 55 & 12 &18 & 30 &  15 & 25 & 55  &13 &24	& 50\\ \hline

\multirow{2}{*}{uDirect}&25 & 40 & 120 & 95 & 140 & 180 &  122  &  155 & 180 & 82 & 121 & 180 \\
&(4.2\%) &	(33.3\%) &	(118.2\%) &(691.7\%) &	(677.8\%) &	(500.0\%) &(713.3\% )&	(520.0\%) &	(227.3\%) &
 (530.8\%)& (404.2\%)& (260.0\%)\\
  \hline
\multirow{2}{*}{Android API} &  57 & 69 & 180 &40 & 49 & 180 & 87 & 157 & 180  &36 & 47 &180 \\

 &		(137.5\%) &	(130.0\%) &	(227.3\%) &	(233.3\%) &	(172.2\%) &	(500.0\%)
&		(480.0\%) &	(528.0\%) &	(227.3\%) &		(176.9\%) &(95.8\%) & (260.0\%)\\
\hline
\multirow{2}{*}{GPS} & 37 & 57 & 177 & 37 & 57 & 177 & 37 & 57 & 177 & 37 & 57 & 177 \\

 &(54.17\%) &	(90.0\%) &	(221.82\%) & (208.3\%) &	(216.67\%) &	(490.0\%)
& (146.67\%) &	(128.0\%) &	(221.82\%) &(184.62\%) &(137.5\%) & (254.0\%)\\

\hline\end{tabular}
}
\caption{Comparison between \sys{}, uDirect\cite{udirect,gluhakdesign}, Android API, and GPS in \emph{\bfseries outdoor} environments. Percentages degradation are calculated relative to \sys{}. GPS results do not depend on the phone position relative to user body.}
\label{tab:comp_out}
\end{table*}

\begin{figure}[!t]
\centering
    \subfigure[Pants pocket\label{subfig-2:p_in}]{%
      \includegraphics[width=0.45\linewidth]{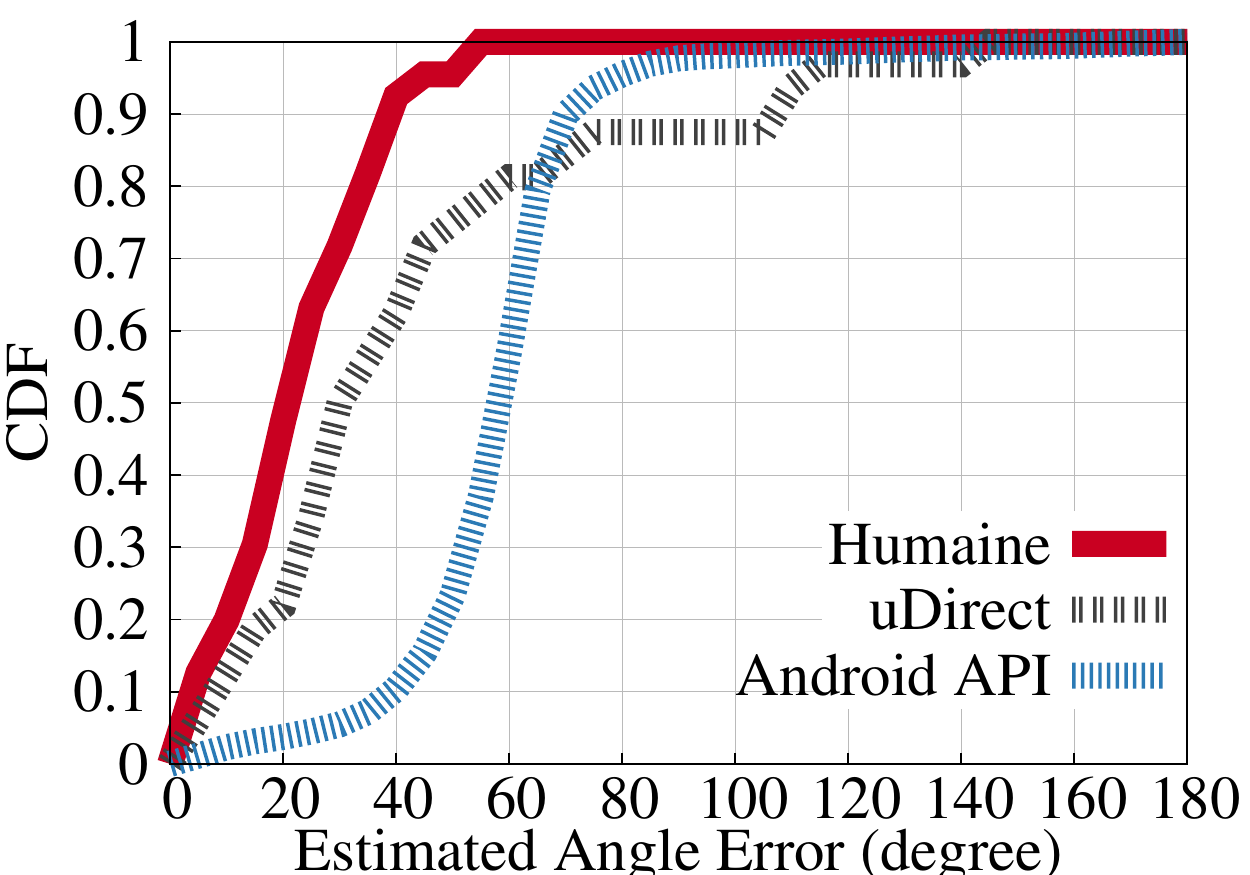}
    }
\hfill
    \subfigure[Hand held\label{subfig-1:h_in}]{
      \includegraphics[width=0.45\linewidth]{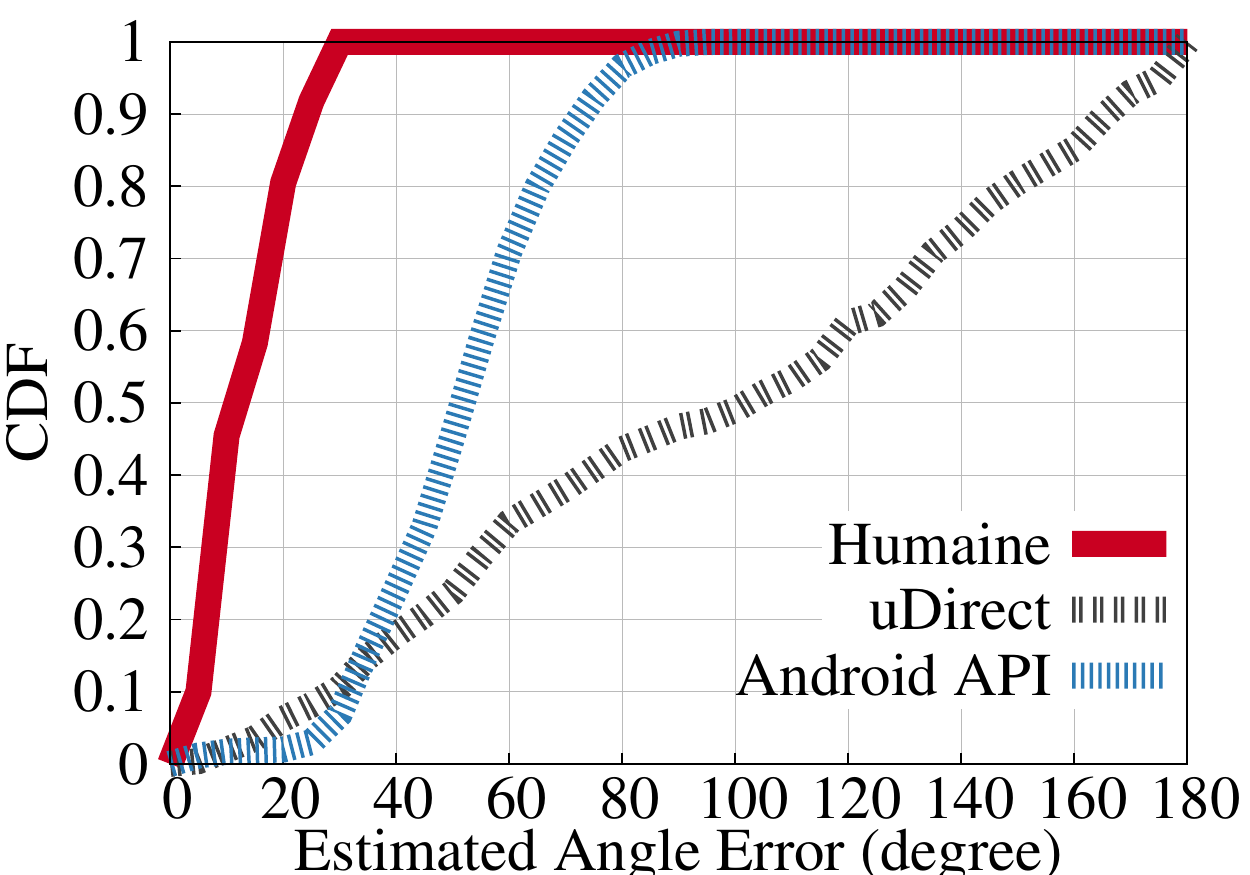}
    }
    \hfill
    \subfigure[Shirt pocket\label{subfig-3:s_in}]{%
      \includegraphics[width=0.45\linewidth]{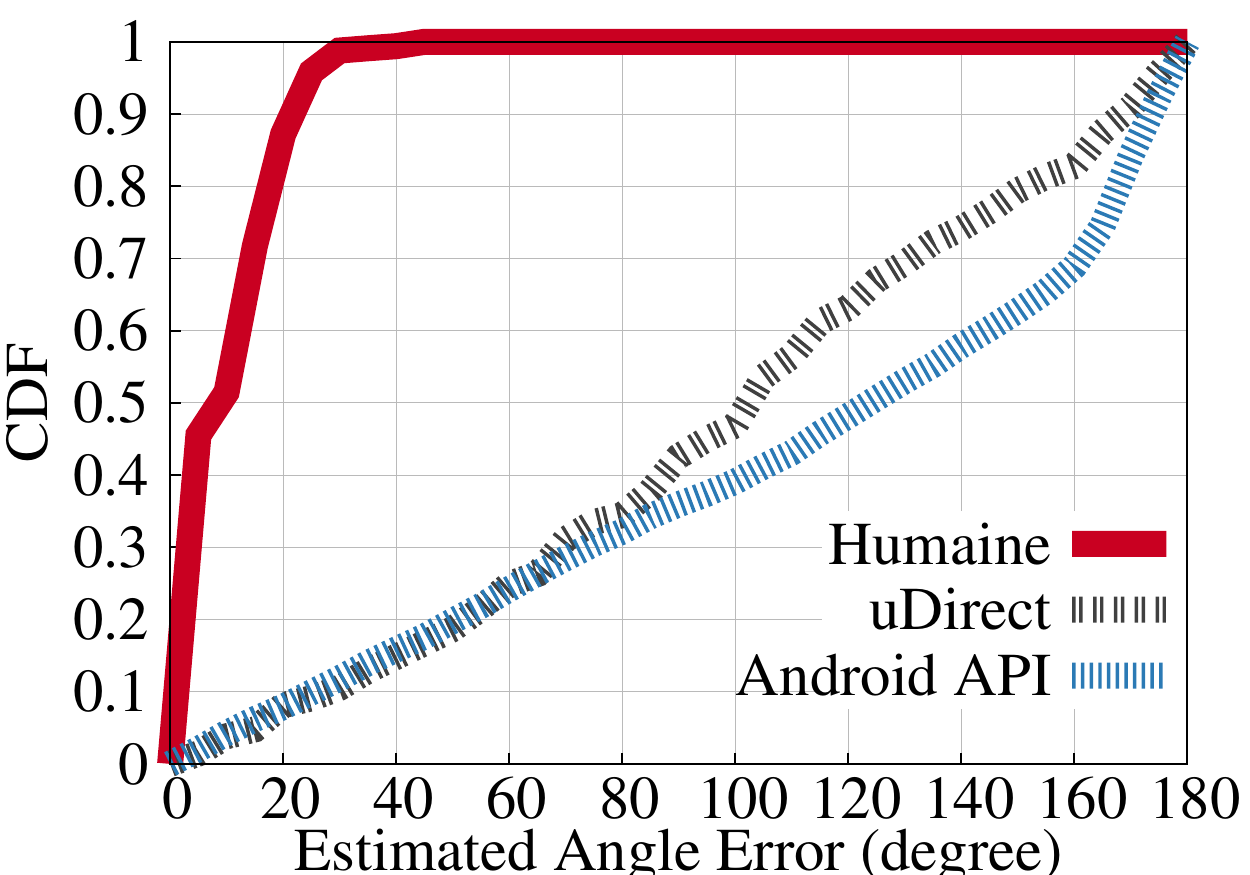}
    }
\hfill
    \subfigure[Bag\label{subfig-3:b_in}]{%
      \includegraphics[width=0.45\linewidth]{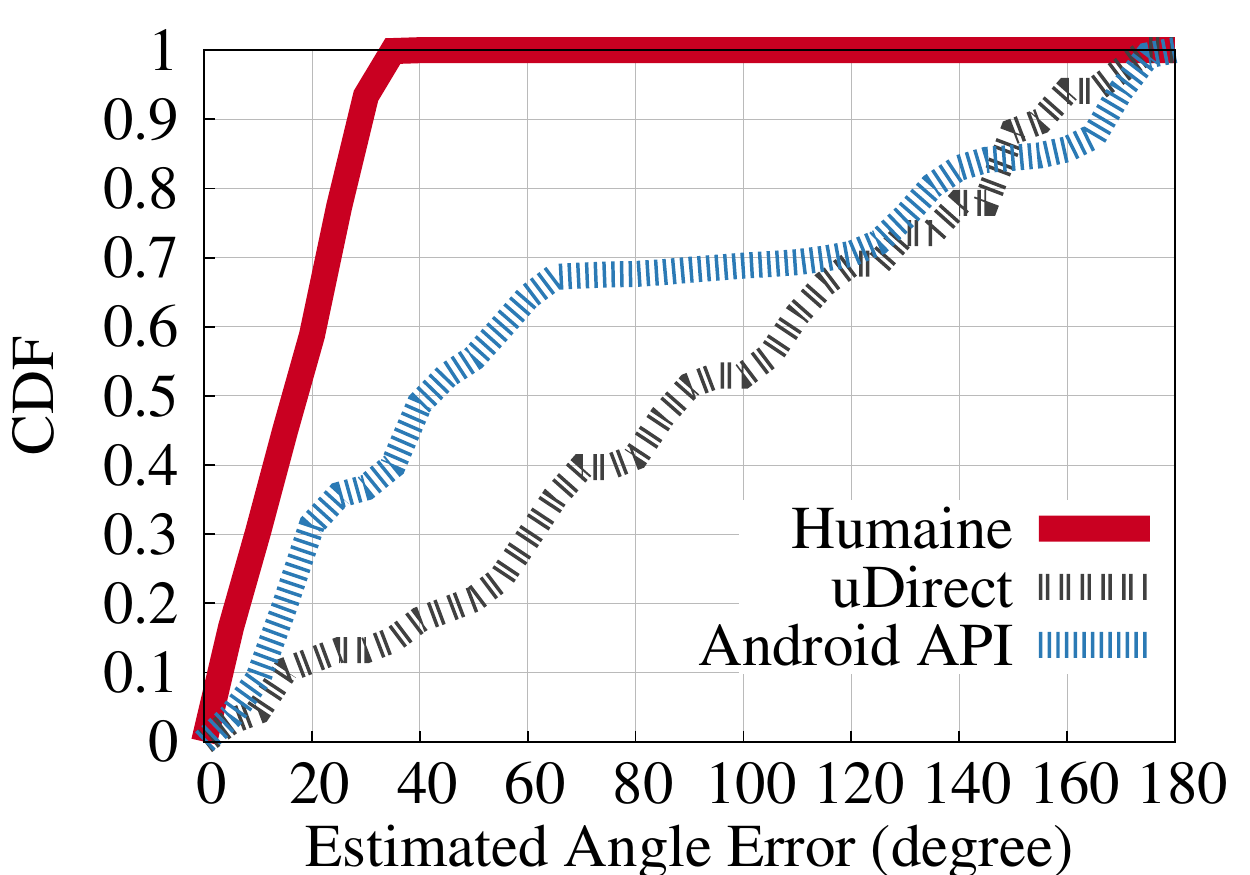}
    }
    \caption{CDF plots comparing \sys{}, uDirect\cite{udirect,gluhakdesign}, and Android API \emph{indoors}. \sys{} outperforms other techniques in all positions. uDirect fails indoors, specially when the phone is not in the pants pocket it was designed for.}
    \label{fig:cdf_in}
  \end{figure}

\begin{figure}[!t]
\centering
\subfigure[Pants pocket\label{subfig-2:p_out}]{%
      \includegraphics[width=0.45\linewidth]{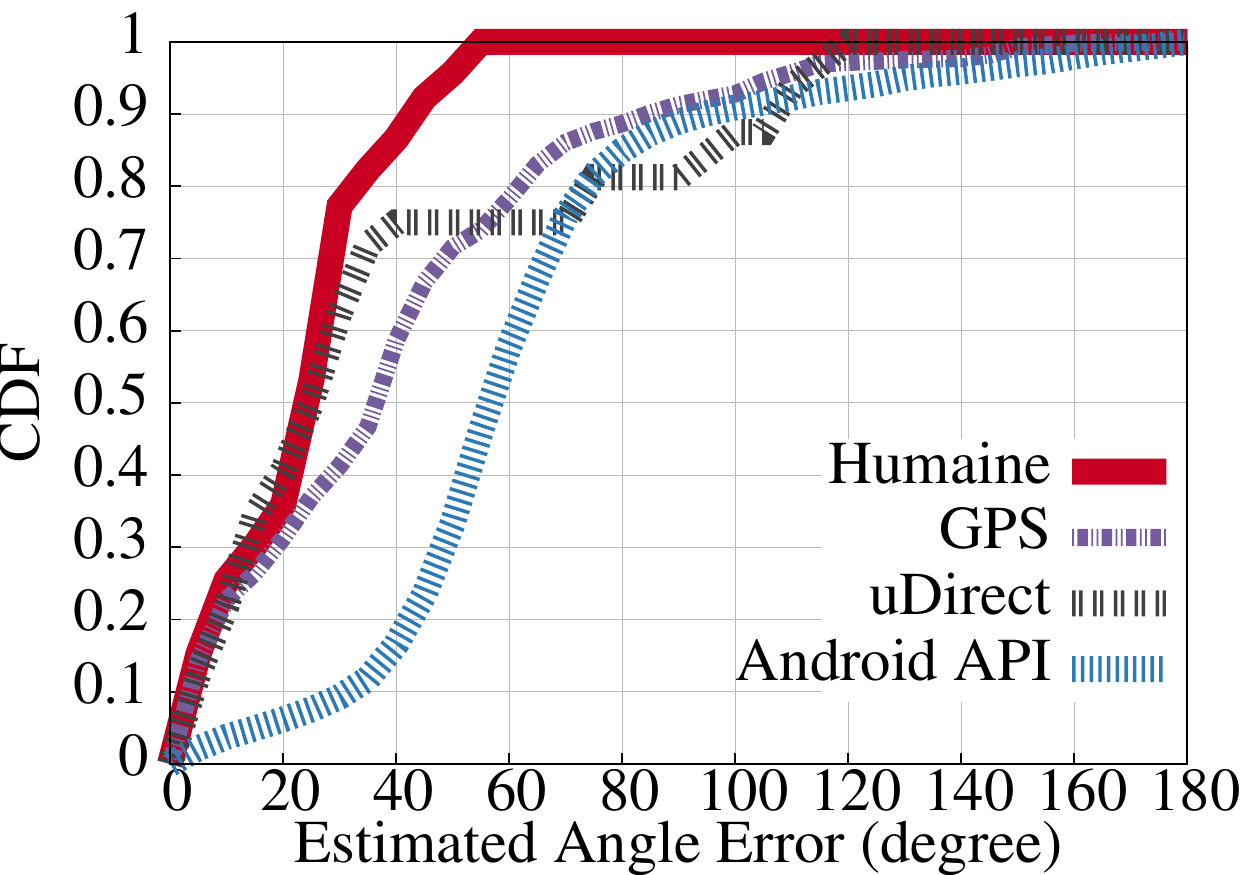}
    }
\hfill
    \subfigure[Hand held]{%
      \includegraphics[width=0.45\linewidth]{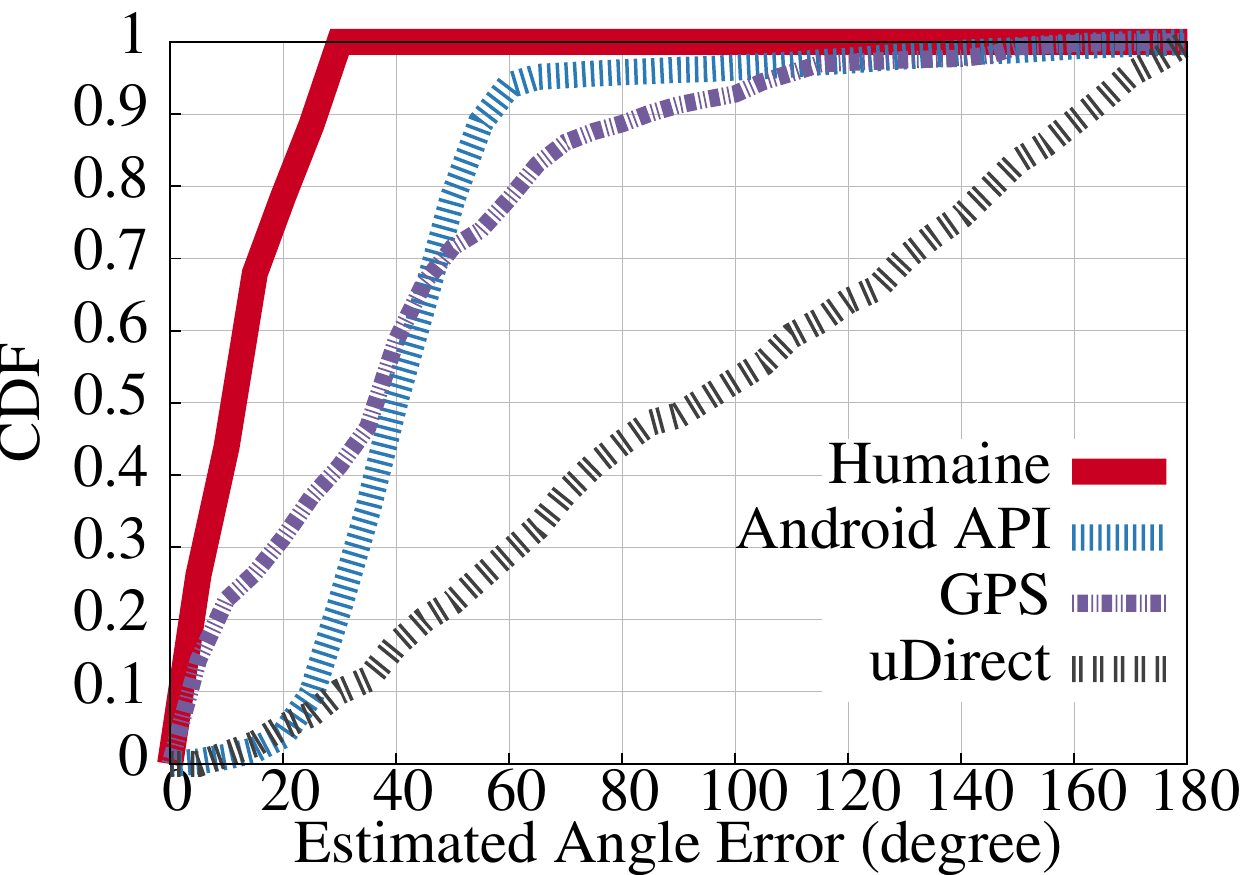}
      \label{subfig-1:h_out}
    }
    \hfill

    \subfigure[Shirt pocket\label{subfig-3:s_out}]{%
      \includegraphics[width=0.45\linewidth]{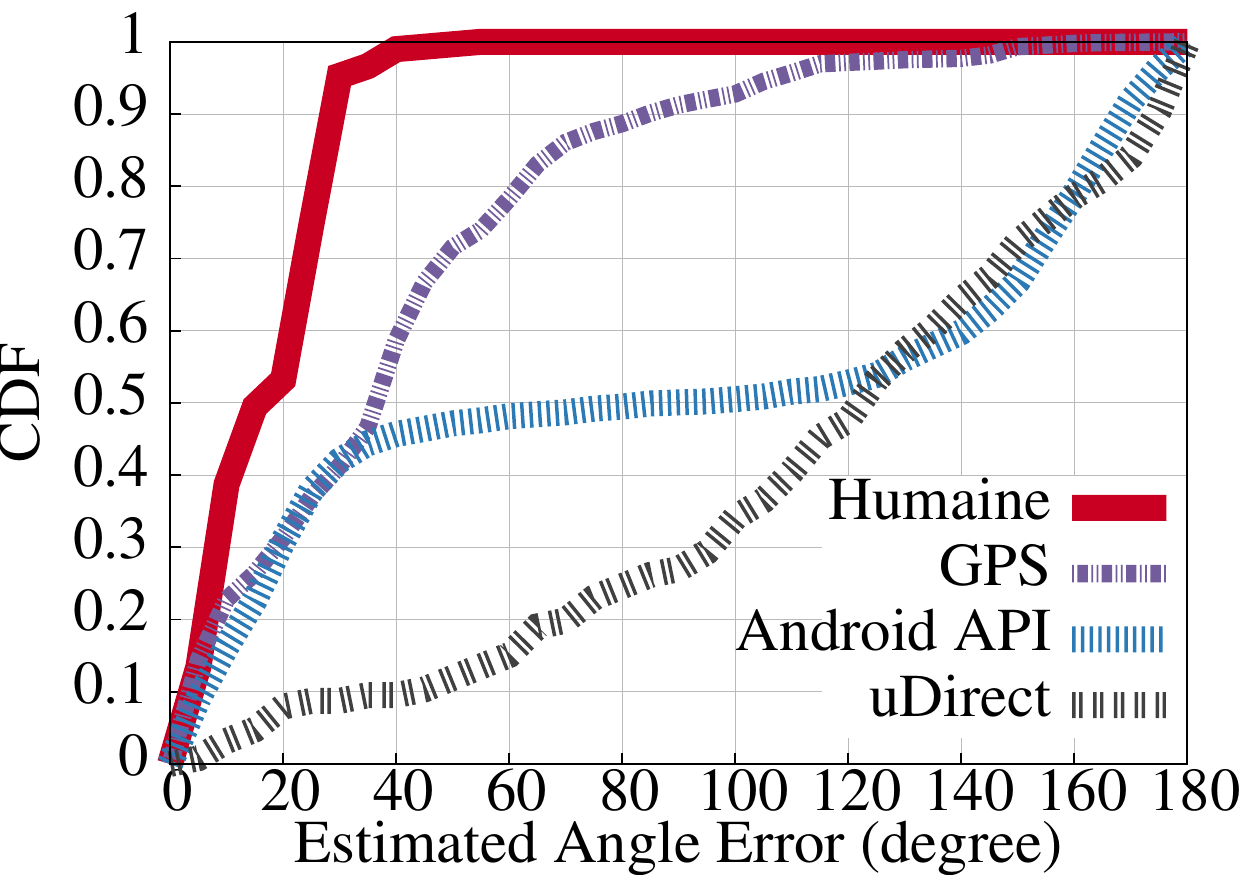}
    }
\hfill
    \subfigure[Bag\label{subfig-3:b_out}]{%
      \includegraphics[width=0.45\linewidth]{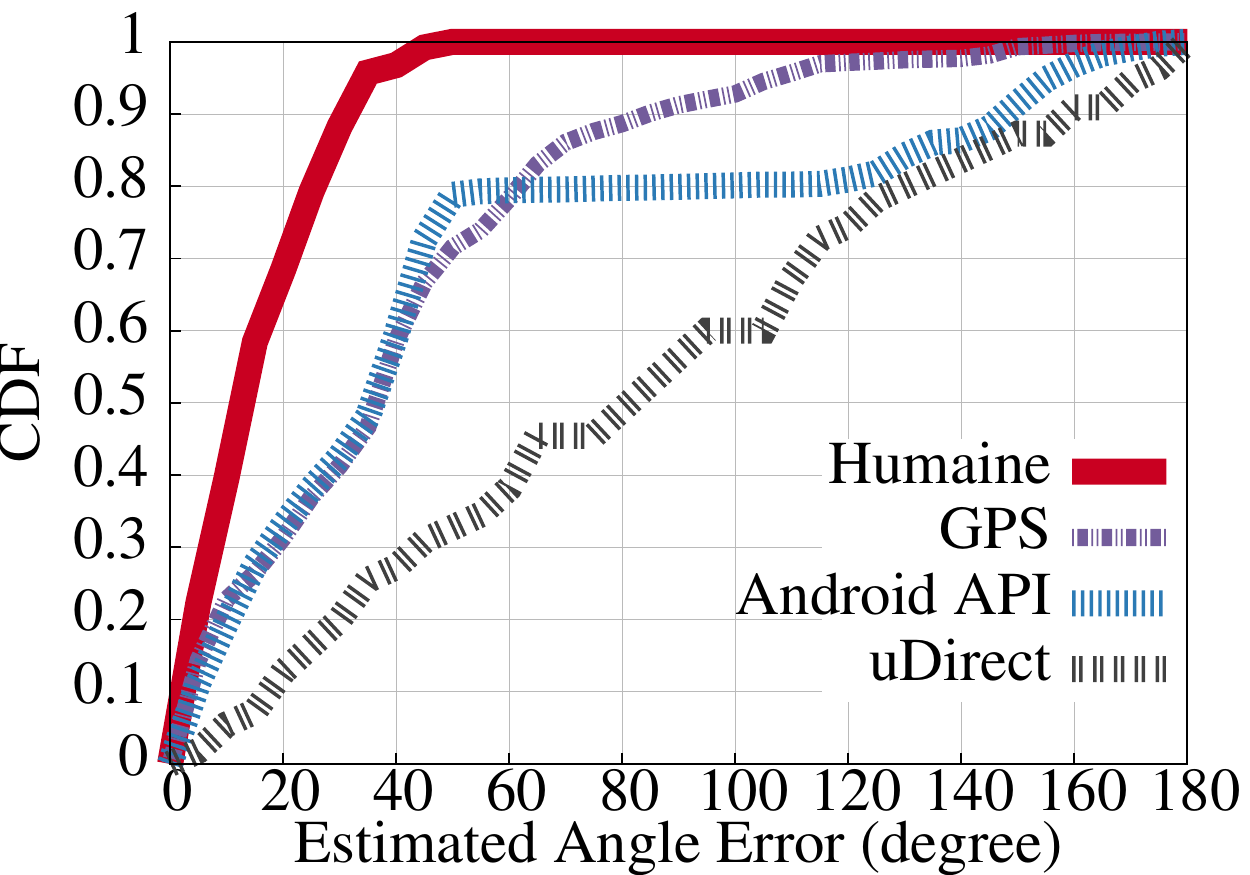}
    }
    \caption{CDF plots comparing \sys{}, GPS, uDirect\cite{udirect,gluhakdesign}, and Android API \emph{outdoors}. \sys{} outperforms all systems in all positions. uDirect fails when the phone is not in the pants pocket (the position it was designed for). GPS gives good accuracy outdoors. However, it has significant latency as we quantify in Section~\ref{sec:res_latency}.}
    \label{fig:cdf_out}
  \end{figure}

In this section, we compare the performance of \sys{}, in terms of heading estimation error, latency in direction change estimation, and power consumption; to uDirect\cite{udirect,gluhakdesign}, the Android API, and GPS. uDirect uses inertial sensors available on the cell phone to estimate the user direction. However, it employs a step detection technique to detect the user heading. uDirect requires a new analytical model of the generated acceleration pattern for each new position~\cite{udirect,gluhakdesign}. However, \textbf{the only model} presented and evaluated in details is for handling the acceleration pattern of the femur bone (i.e. in pants pocket position), and deriving models for other positions is not straight forward as we show later in this section. On the other hand, {\bf the Android API is the default heading estimation technique available on Android phones and has been used by a number of systems, e.g. \cite{constandache2010compacc,youssef2010gac}, to determine the \emph{phone orientation}, rather than the human orientation (assuming both are the same)}. 
 However, it is considered \textbf{\emph{the widely deployed}} state-of-the-art. We also compare with the user bearing provided by the cell phone's GPS as widely used \textbf{\emph{outdoor}} technique that can estimate the user heading and does not depend on the phone position.

We start by evaluating the accuracy of the different techniques on a \textbf{single} segment to quantify the effect of different phone positions. We then perform a continuous trace experiment to study the latency of the different techniques and their suitability for real-time applications.
\subsection{One Segment User's Heading Estimation Error}

Figures \ref{fig:cdf_in} and \ref{fig:cdf_out} show the CDFs of the user orientation estimation error for the three systems---\sys{}, uDirect\cite{udirect,gluhakdesign}, and the Android API---indoors and outdoors respectively for the different cell phone positions under 17 different testbeds. 
 The experiments were performed by 10 users while placing the phones at different positions covering different phone orientations relative to the user direction of motion. We also show the GPS for outdoor experiments. Tables \ref{tab:comp_in} and \ref{tab:comp_out} summarize the results. Through this evaluation, we used straight segments, i.e. \textbf{without including any changes in direction.} We evaluate the effect of the change in direction on the techniques estimation error and latency in Section~\ref{sec:res_latency}.

\begin{figure}[!t]
\centering
\includegraphics[width=0.95\linewidth]{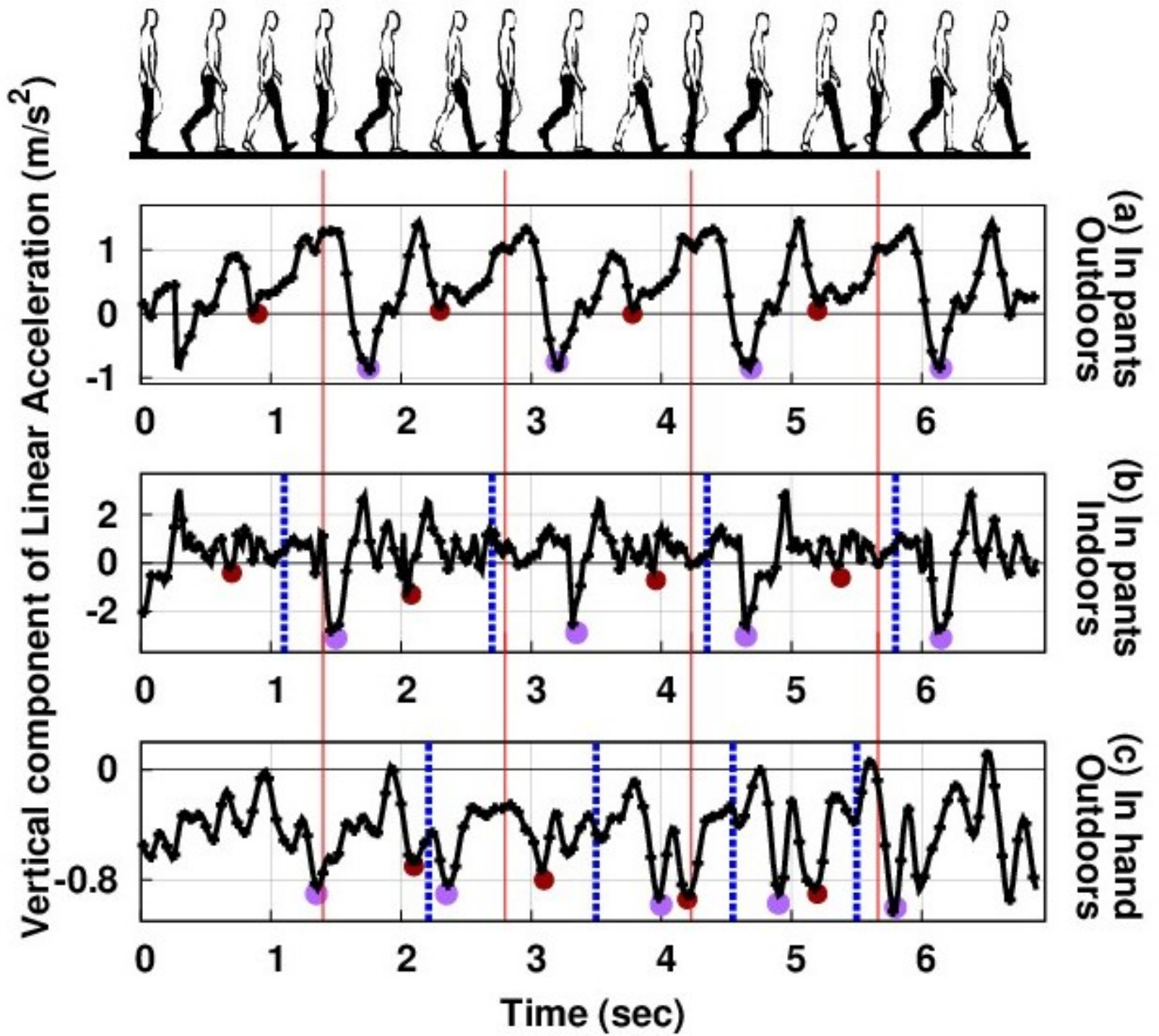}
\caption{The vertical acceleration used by uDirect \cite{udirect,gluhakdesign} for estimating the user heading (solid red lines represent the ground truth, circles represent the global and local minimums at each step used to estimate the point of foot impact, dotted blue lines represent the estimated point by uDirect): (a) in pants pocket outdoors (best), (b) in pants pocket indoors and (c) hand held outdoors.  Due to magnetic noise indoors, the step pattern is deformed; explaining uDirect performance degradation indoors. Also, the in-pocket step model cannot be used directly with other positions (e.g. shown in-hand). }
\label{fig:udirect}
\end{figure}

The results show that \sys{} gives the best accuracy in all cases. Although both \sys{} and uDirect use the magnetometer sensor, which is sensitive to magnetic noise in the environment especially indoors; we believe that \sys{} is less sensitive due to two reasons: 
 First \sys{} depends on estimating the heading direction based on the \emph{variance} of the user linear acceleration, which is less sensitive to noise compared to the rotated acceleration \emph{pattern} within a step used by uDirect. Second, \sys{} sensor fusion module described in Section~\ref{sec:fusion} lessens the effect of the magnetometer noise or drift. For example, Figure~\ref{fig:udirect} shows the user step pattern used by uDirect in indoor and outdoor environments. The figure shows that the step pattern is noisy in indoor environment, significantly affecting uDirect accuracy.

 The same figure also show that uDirect is sensitive to the movement model used. It has been designed to work with the in-pocket position, which gives the best accuracy for it. However using this model with other phone positions leads to excessive errors.

 The Android standard API, as it does not estimate the user heading, has large errors, especially in the shirt pocket position. At this position, the cell phone is vertically orientated relative to the user's direction, leading to the maximum error.

GPS has good accuracy outdoors when the user is \textbf{\emph{not changing her direction}}. However, it has a large fix time (average two minutes), excessive latency when the user makes a turn, and its accuracy in urban area decreases significantly. This is quantified in the next section.

\subsection{Performance in a Continuous Trace}
\label{sec:res_latency}
\begin{figure}[!t]
\centering
\subfigure[Motion trajectory]{
\includegraphics[width=0.8\linewidth]{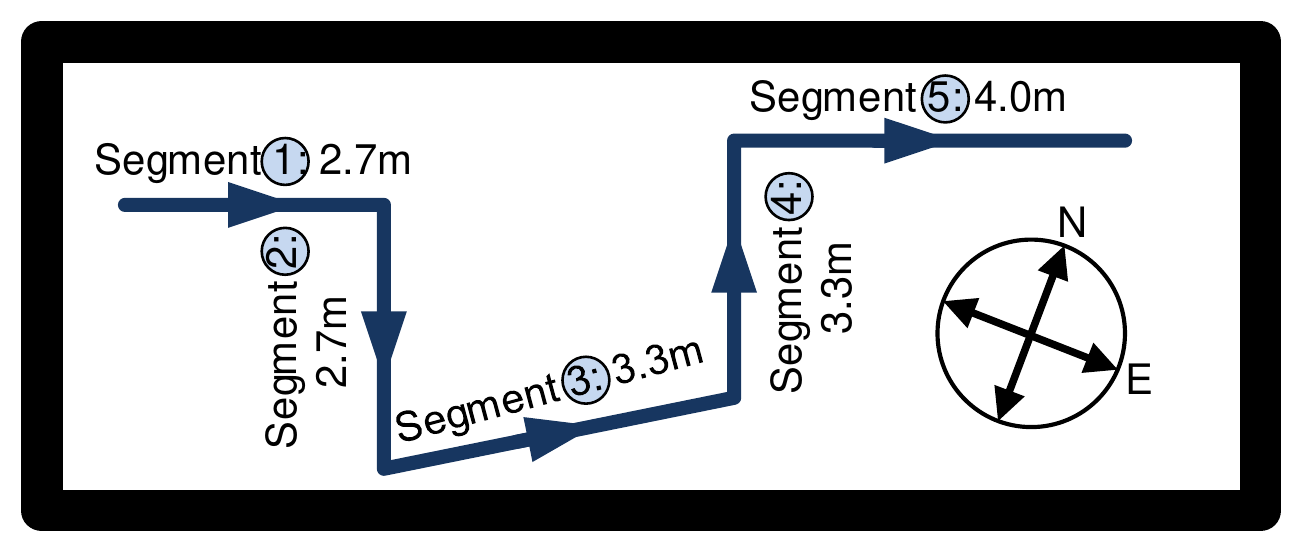}}
\subfigure[Motion tracking]{
\includegraphics[width=0.85\linewidth]{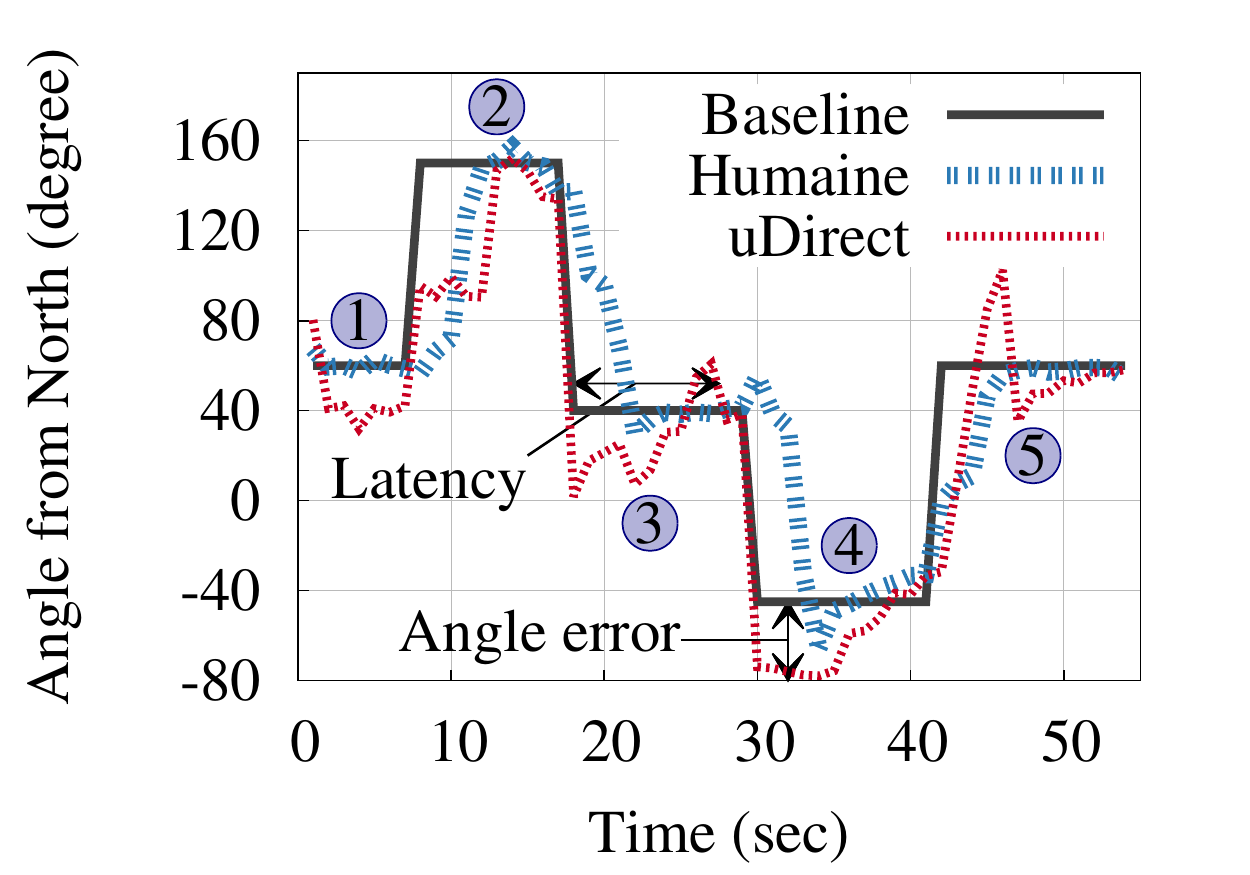}}
\caption{Tracking performance for \sys{} and uDirect\cite{udirect,gluhakdesign} for a continuous motion trace in an indoor environment. Phone placed \textbf{in pants pocket only} as uDirect completely fails in other positions as in Figure~\ref{fig:cdf_in}.}
\label{fig:trajectory_in}
\end{figure}

\begin{figure}[!t]
\centering
\subfigure[Motion trajectory]{
\includegraphics[width=0.8\linewidth]{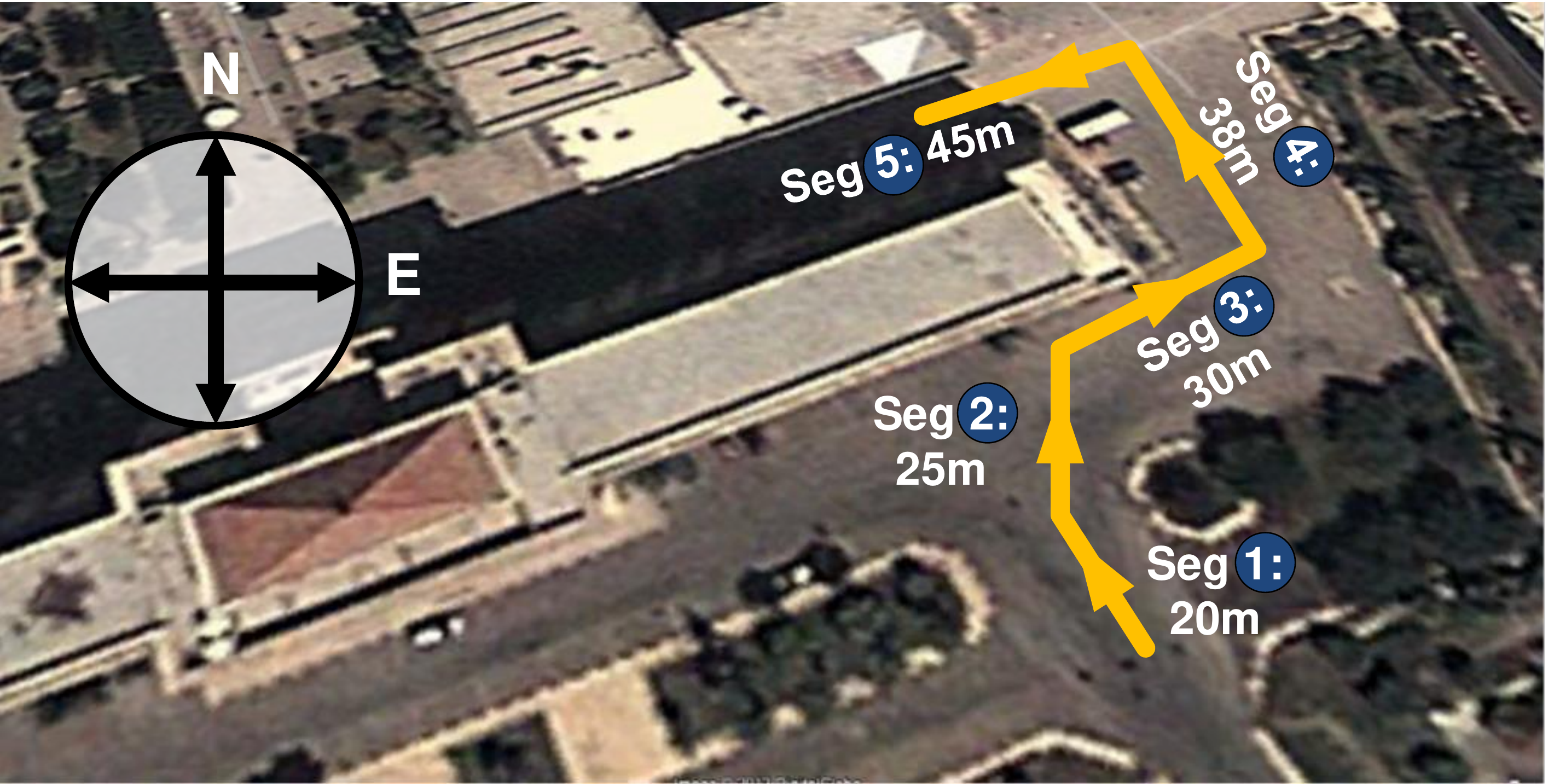}}
\subfigure[Motion tracking]{
\includegraphics[width=0.85\linewidth]{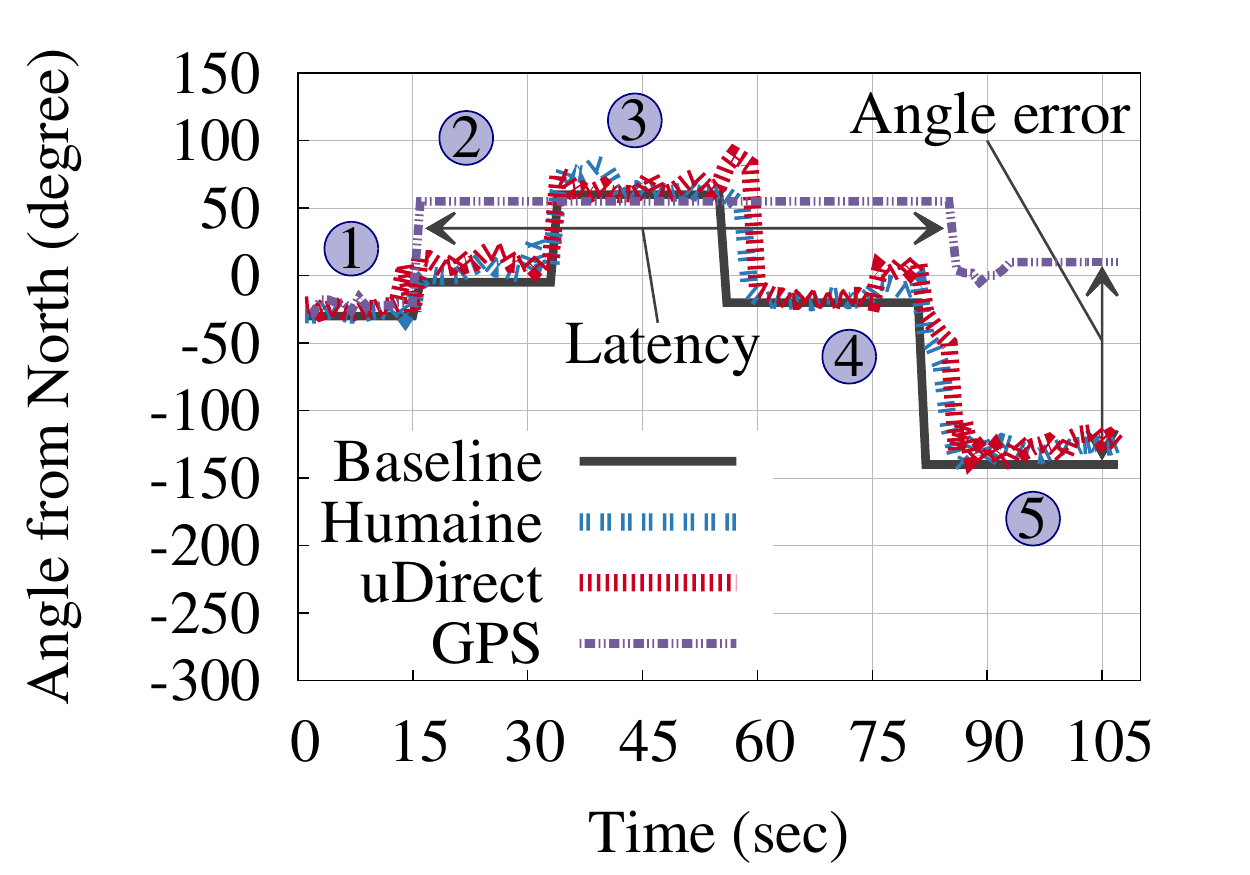}}
\caption{Tracking performance for \sys{}, uDirect\cite{udirect,gluhakdesign}, and GPS for a continuous motion trace in an outdoor environment. \textit{Note that GPS can track the angle accurately within a \textbf{single} segment but gives significant latency (70 seconds on average)}. Phone placed \textbf{in pants pocket only} as uDirect completely fails in other positions as in Figure~\ref{fig:cdf_out}.}
\label{fig:trajectory_out}
\end{figure}

Figures \ref{fig:trajectory_in} and \ref{fig:trajectory_out} show the trajectory of the experiments to examine the effect of changing the direction during a continuous trace on \sys{} and uDirect\cite{udirect,gluhakdesign} indoors and on \sys{}, uDirect\cite{udirect,gluhakdesign} and GPS outdoors. Table~\ref{tab:sum_traj} summarizes the results. For both cases, the phone was placed in the pants pocket \emph{as uDirect does not perform well in other cases, i.e. \textbf{this is the best case scenario for uDirect}} as shown in the previous section. The results confirm that \sys{} can smoothly track the user heading in both indoors and outdoors environments with most of the errors at the instances of direction change. \sys{} also has the minimum latency, highlighting its suitability for applications that can tolerate delay. 

 In addition, uDirect accuracy is significantly affected in indoor environments due to the noise affecting the acceleration pattern used in heading estimation (Figure~\ref{fig:udirect}).

GPS suffers from degradation in accuracy at areas without clear view of the sky, e.g. indoors, and urban canyons~\cite{aly2013dejavu}. In addition, GPS has an initial fix latency with an average over different devices of 2 minutes. Moreover, it suffers from \textbf{\emph{latency when changing direction}} with an average of 70 seconds. This latency affects the performance of GPS as shown in the experiment in Figure~\ref{fig:trajectory_out}.

\begin{table*}[!t]
\centering
\resizebox{\textwidth}{!}{\small
\begin{tabular}{|c|c|c|c||c||c|c|c||c|||c|}\hline \multirow{3}{*}{Environment} & \multicolumn{4}{c||}{Indoors} & \multicolumn{4}{c||}{Outdoors} &\multirow{3}{*}{Power(mW)}\\\cline{2-9} & \multicolumn{3}{c|}{Abs. Error (degree)} & Latency (sec) & \multicolumn{3}{c|}{Abs. Error (degree)} & Latency (sec)&\\\cline{2-9} &Min & Average & Max & Average &Min & Average & Max & Average  &\\\hline \hline Humaine & 0.022 & 16.24 & 83.71 &3.5 & 0.007 & 11.86 & 95.13& 3 & 125.5\\\hline
\multirow{2}{*}{uDirect} & 1.763 & 22.49 & 91.69 &5.75 & 0.111 & 17.41& 111.17 &4 & 113.5\\
&(7913.6\%) & (38.5\%) & (9.5\%) & (64.3\%) & (1485.7\%) & (46.8\%) & (16.9\%) & (33.3\%) & (-9.5\%)  \\\hline
\multirow{2}{*}{GPS} &  &  &  & & 2 & 66.94& 165 &70 & 315\\
& - & - & - & - & (28471.4\%) & (464.4\%) & (73.45\%) & (2233.3\%)  & (150.9\%) \\\hline
\end{tabular}
}
\caption{Summary for the continuous-trace experiments. Phone placed \textbf{only in pants pocket} as uDirect completely fails in other positions as in Figure~\ref{fig:cdf_in}. Percentages degradation are calculated relative to \sys{}. The table shows that \sys{} outperforms uDirect algorithm in both accuracy and latency with comparable energy-consumption.}
 \label{tab:sum_traj}
 \end{table*}

\subsection{Power Consumption}

\begin{figure}[!t]
\centering
\includegraphics[width=0.7\linewidth]{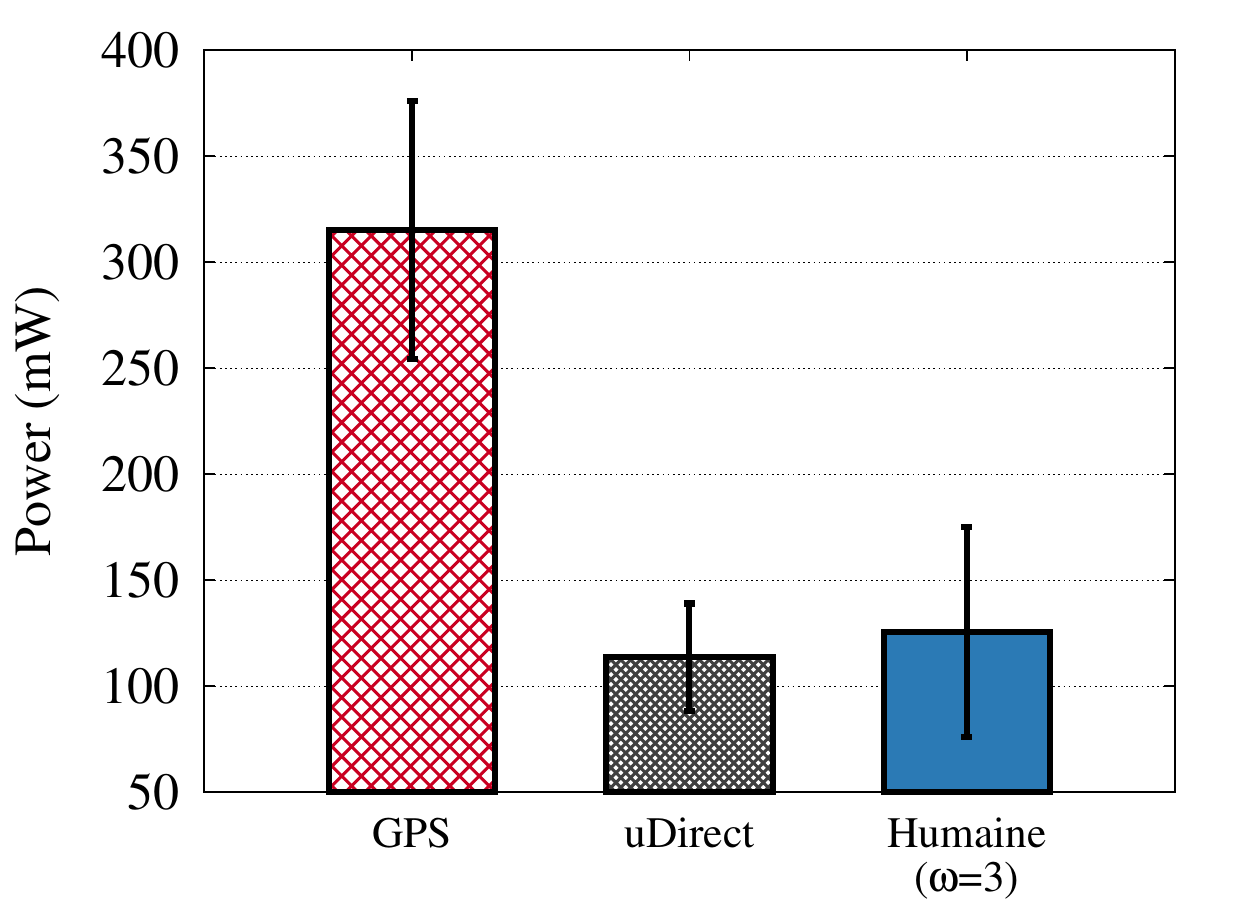}
\caption{Power consumption for the different algorithms.}
\label{fig:power}
\end{figure}
Figure~\ref{fig:power} and Table~\ref{tab:sum_traj}  summarize the power consumption of the different algorithms. To calculate the power consumption, we used the PowerTutor profiler\cite{zhang2010accurate}. The figure shows that as expected, GPS has the worst energy consumption. uDirect involves searching for a specific pattern inside a step. Therefore, its complexity is linear in the step length. 
 The results shows that \sys{} energy consumption is comparable to uDirect. This can be further enhanced by other techniques, e.g. duty-cycling.  
\section{Conclusion}
\label{sec:conc}
We introduced \sys{}, a system for robustly estimating the orientation of users carrying cell phones, suitable for new emerging crowd-sensing applications where the user has the freedom to have her phone in any arbitrary position or orientation. It fuses the phone's energy-efficient inertial sensors and applies the PCA technique effectively to detect the user direction.

Evaluation of \sys{} on Android devices with {\it \bfseries 170 experiments} on different testbeds and phone positions shows that the system works accurately indoors and outdoors, without user involvement or prior configurations. \sys{} achieved a median accuracy of  $15^\circ$; which is $558\%$ better than the state-of-the-art. 
Furthermore, \sys{} has a minimal effect on the phone battery and latency, highlighting its suitability for real-time applications.

Currently, we are extending the system in different directions including employing other sensors available on the cell phone, e.g. the camera, opportunistically for a more accurate direction estimation, duty-cycling the sensors for further energy efficiency, dynamically adapting the system parameters, among others.

\section*{Acknowledgment}
This work was supported in part by a Google Research Award.



%
{
\bibliographystyle{abbrv}
\bibliography{user-orient}

\begin{thebibliography}{10}

\bibitem{aly_map14}
H.~Aly, A.~Basalamah, and M.~Youssef.
\newblock Map++: {A} crowd-sensing system for automatic map semantics
  identification.
\newblock In {\em SECON}. IEEE, 2014.

\bibitem{aly2013dejavu}
H.~Aly and M.~Youssef.
\newblock Dejavu: an accurate energy-efficient outdoor localization system.
\newblock In {\em SIGSPATIAL}. ACM, 2013.

\bibitem{alzantot2012crowdinside}
M.~Alzantot and M.~Youssef.
\newblock {CrowdInside}: Automatic construction of indoor floorplans.
\newblock In {\em ACM SIGSPATIAL GIS}, 2012.

\bibitem{alzantot2012uptime}
M.~Alzantot and M.~Youssef.
\newblock {UPTIME}: Ubiquitous pedestrian tracking using mobile phones.
\newblock In {\em WCNC}. IEEE, 2012.

\bibitem{omnidirectional}
S.~Beauregard.
\newblock Omnidirectional pedestrian navigation for first responders.
\newblock In {\em Proc. of Workshop on Pos., Navig. and Comm.} IEEE, 2007.

\bibitem{blum2012smartphone}
J.~R. Blum, D.~Greencorn, and J.~R. Cooperstock.
\newblock Smartphone sensor reliability for augmented reality applications.
\newblock In {\em MobiQuitous}, 2012.

\bibitem{constandache2010compacc}
I.~Constandache, R.~R. Choudhury, and I.~Rhee.
\newblock Compacc: Using mobile phone compasses and accelerometers for
  localization.
\newblock {\em INFOCOM}, 2010.

\bibitem{constandache2010towards}
I.~Constandache, R.~R. Choudhury, and I.~Rhee.
\newblock Towards mobile phone localization without war-driving.
\newblock In {\em INFOCOM}. IEEE, 2010.

\bibitem{gpsuc}
Y.~Cui and S.~S. Ge.
\newblock Autonomous vehicle positioning with {GPS} in urban canyon
  environments.
\newblock {\em IEEE Trans. on Robotics and Automation}, 2003.

\bibitem{goldstein1980classical}
H.~Goldstein.
\newblock {\em Classical mechanics}.
\newblock Addison-Wesley, 1980.

\bibitem{hamilton1866elements}
W.~Hamilton and W.~Hamilton.
\newblock {\em Elements of Quaternions}.
\newblock Longmans, Green, \& Company, 1866.

\bibitem{gps}
B.~Hofmann-Wellenhof, H.~Lichtenegger, and J.~Collins.
\newblock {\em Global Positioning System: Theory and Practice}.
\newblock Springer, 1993.

\bibitem{udirect}
S.~A. Hoseinitabatabaei, A.~Gluhak, and R.~Tafazolli.
\newblock udirect: A novel approach for pervasive observation of user direction
  with mobile phones.
\newblock In {\em PerCom}. IEEE, 2011.

\bibitem{gluhakdesign}
S.~A. Hoseinitabatabaei, A.~Gluhak, R.~Tafazolli, and W.~Headley.
\newblock Design, realization, and evaluation of {uDirect}--an approach for
  pervasive observation of user facing direction on mobile phones.
\newblock {\em IEEE TMC}, 2014.

\bibitem{kourogi2003method}
M.~Kourogi and T.~Kurata.
\newblock A method of personal positioning based on sensor data fusion of
  wearable camera and self-contained sensors.
\newblock In {\em IEEE Inter. Conf. on Multisensor Fusion and Integration for
  Intelligent Sys.}, 2003.

\bibitem{kourogi2003personal}
M.~Kourogi and T.~Kurata.
\newblock Personal positioning based on walking locomotion analysis with
  self-contained sensors and a wearable camera.
\newblock In {\em Inter. Symp. on Mixed and Augmented Reality}. IEEE, 2003.

\bibitem{kourogi2003wearable}
M.~Kourogi and T.~Kuratta.
\newblock A wearable augmented reality system with personal positioning based
  on walking locomotion analysis.
\newblock In {\em ISMAR}. IEEE, 2003.

\bibitem{pca1}
K.~Kunze, P.~Lukowicz, K.~Partridge, and B.~Begole.
\newblock Which way am {I} facing: Inferring horizontal device orientation from
  an accelerometer signal.
\newblock In {\em Inter. Symp. on Wearable Comps}. IEEE, 2009.

\bibitem{li2012reliable}
F.~Li, C.~Zhao, G.~Ding, J.~Gong, C.~Liu, and F.~Zhao.
\newblock A reliable and accurate indoor localization method using phone
  inertial sensors.
\newblock In {\em Ubicomp}. ACM, 2012.

\bibitem{ronald1998magnetic}
R.~T. Merrill, W.~McElhinny, and P.~McFadden.
\newblock {\em The Magnetic Field of the Earth: Paleomagnetism, the Core, and
  the Deep Mantle}.

\bibitem{schall2013mobile}
G.~Schall.
\newblock {\em Mobile Computing: Mobile Augmented Reality for Human Scale
  Interaction with Geospatial Models: the Benefit for Industrial Applications}.
\newblock Springer, 2013.

\bibitem{pca2}
U.~Steinhoff and B.~Schiele.
\newblock Dead reckoning from the pocket-an experimental study.
\newblock In {\em PerCom}. IEEE, 2010.

\bibitem{headio}
Z.~Sun, S.~Pan, Y.-C. Su, and P.~Zhang.
\newblock Headio: zero-configured heading acquisition for indoor mobile devices
  through multimodal context sensing.
\newblock UbiComp '13. ACM.

\bibitem{wang2012no}
H.~Wang, S.~Sen, A.~Elgohary, M.~Farid, M.~Youssef, and R.~R. Choudhury.
\newblock No need to war-drive: Unsupervised indoor localization.
\newblock In {\em MobiSys}. ACM, 2012.

\bibitem{wang2010novel}
Q.~Wang, X.~Zhang, X.~Chen, R.~Chen, W.~Chen, and Y.~Chen.
\newblock A novel pedestrian dead reckoning algorithm using wearable emg
  sensors to measure walking strides.
\newblock In {\em UPINLBS, 2010}. IEEE.

\bibitem{youssef2010gac}
M.~Youssef, M.~A. Yosef, and M.~El-Derini.
\newblock {GAC}: Energy-efficient hybrid {GPS}-accelerometer-compass {GSM}
  localization.
\newblock In {\em GLOBECOM}. IEEE, 2010.

\bibitem{zhang2010accurate}
L.~Zhang, B.~Tiwana, Z.~Qian, Z.~Wang, R.~P. Dick, Z.~M. Mao, and L.~Yang.
\newblock Accurate online power estimation and automatic battery behavior based
  power model generation for smartphones.
\newblock In {\em IEEE/ACM/IFIP Inter conf. CODES+ISSS}, 2010.

\bibitem{zhang2012motion}
M.~Zhang and A.~A. Sawchuk.
\newblock Motion primitive-based human activity recognition using a
  bag-of-features approach.
\newblock In {\em ACM SIGHIT Inter. Health Inform. Symp.}, 2012.

\end{thebibliography}
}

\end{document}